\documentstyle[epsfig,12pt]{article}

\begin{document}

\title{Low-energy spectrum of $SU(3)$ Yang-Mills Quantum Mechanics}
\author{Hans-Peter Pavel \\[1cm]
Bogoliubov Laboratory of Theoretical Physics,
\\
Joint Institute for Nuclear Research, Dubna, Russia\footnote{email: pavel@theor.jinr.ru}
}
\date{December 12th, 2021}
\maketitle

\begin{abstract}
The SU(3) Yang-Mills Quantum Mechanics of spatially constant gluon fields is considered in the 
unconstrained Hamiltonian approach using  the "flux-tube gauge".
The Faddeev-Popov operator, its determinant and inverse, are rather simple, but 
show a highly non-trivial periodic structure of six Gribov-horizons separating six Weyl-chambers.
The low-energy eigensystem of the obtained physical Hamiltonian can be calculated (in principle with arbitrary high precision) 
using the orthonormal basis of eigenstates of the corresponding harmonic oscillator problem with the same non-trivial Jacobian
only replacing the chromomagnetic potential by the 16dimensional harmonic oscillator potential.
This turns out to be integrable and its eigenstates  be made out of orthogonal polynomials of the 45 components of eight irreducible
symmetric tensors.
The calculations in this work have been carried out in all sectors $J^{PC}$ up to spin $J=11$, and up to polynomial order
$10$ for even and $11$ for odd parity. 
The low-energy eigensystem of the physical Hamiltonian of $SU(3)$ Yang-Mills Quantum Mechanics is found to converge nicely 
when truncating at higher and higher polynomial order (equivalent to increasing the resolution in functional space).
Our results are in good agreement with the results of Weisz and Zieman (1986) using the constrained Hamiltonian approach.
We find excellent agreement in the $0^{++}$ and $2^{++}$ sectors, much more accurate values in other sectors
considered by them, e.g. in the $1^{--}$ and $3^{--}$ sectors , and quite accurate "new results" for the sectors not considered by them,
e.g. $2^{--}, 4^{--}, 5^{--}, 3^{++}$.
\end{abstract}


\section{Introduction}

The Yang-Mills  Quantum Mechanics (YM QM) of spatially constant gluon fields has been studied for a long time,
as a toy model for the QCD vacuum\cite{BasMatSav}-\cite{Martin}, 
as the zeroth order of a weak coupling expansion \cite{Luescher}-\cite{Weisz and Ziemann},
 and as zeroth order of a strong coupling expansion \cite{KP1}-\cite{pavel2010}. 
For the case of SU(2) YM theory, the symmetric gauge turned out to exist in the strong coupling limit and to be very convenient for 
calculations, also including the quarks \cite{pavel2011}. The reduced gauge fields transform as symmetric tensors under spatial rotations, 
and the Faddeev-Popov (FP) operator turned out to be non-trivial but managable in a way similar to the Calogero model \cite{Calogero}. 
For the case of SU(3) YM theory, however, the symmetric
gauge can also be defined \cite{pavel2012}-\cite{pavel2014} and leads to reduced fields transforming as tensors 
under spatial rotations, but the corresponding FP-operator turns out to be very complicated.
In previous work \cite{pavel2016}, a new algebraic gauge for SU(3) YM theory, the flux-tube gauge, has been proposed,
which exists in the strong coupling limit and has a simple non-trivial FP-operator. As for the case of SU(2) YM theory
in the symmetric gauge \cite{pavel2010}, the corresponding gauge reduced SU(3) YM
Hamiltonian in the flux-tube gauge can be expanded in strong coupling $\lambda=g^{-2/3}$ with the leading order corresponding to  
SU(3) YM QM of spatially constant fields. 
The drawback of the flux-tube gauge is however, that the reduced fields $A$ are color-singlets, 
but are not transforming as tensors under spatial rotations.
We shall show in this article how this 
can be circumvened by forming certain irreducible polynomials of the reduced $A$, symmetric tensors
with definite eigenvalues of J,P, and C.
We shall calculate in this work the low-energy spectrum of SU(3) YM QM  in the 
unconstrained Hamiltonian approach using the flux-tube gauge.
The Faddeev-Popov operator, its determinant and inverse, are rather simple, but 
show a highly non-trivial periodic structure of six Gribov-horizons separating six Weyl-chambers.
The low-energy eigensystem of the obtained physical Hamiltonian can be calculated (in principle with arbitrary high precision) 
using the orthonormal basis of eigenstates of the corresponding harmonic oscillator problem with the same non-trivial Jacobian
only replacing the chromomagnetic potential by the 16-dimensional harmonic oscillator potential. 
This turns out to be integrable and its eigenstates to be made out of orthogonal polynomials of the 45 components of eight elementary
spatial tensors. 

The paper is organised as follows. In Section 2 we give a short introduction to the Hamiltonian approach of
SU(3) Yang-Mills Quantum mechanics of spatially constant gluon fields $V_{ai}$ constrained by non-Abelian Gauss laws.
 In Section 3 the flux-tube gauge is defined and
shown to lead to a rather simple but non-trivial Faddeev-Popov (FP) operator. In Section 4 the corresponding harmonic
oscillator (HO) Hamiltonian, obtained by replacing the chromomagnetic potential by an 16-dimensional harmonic oscillator
potentials, is solved analytically by polynomials in the components of eight irreducible tensors in reduced space $A$. 
In Section 5 we use the obtained eigensystem of the HO-Hamiltonian to find the eigensytem of the Hamiltonian of
SU(3)-Yang-Mills QM in dependence of truncation at higher and higher polynomial degree. 
The results are compared with those of Weisz and Zieman \cite{Weisz and Ziemann} obtained in the constrained approach,
and comparison with the low glueball spectrum obtained in \cite{Morningstar},\cite{Chen} using lattice QCD.
Section 6 gives our Conclusions. Some technical
details are banned to Appendices A-F.


\section{Constrained SU(3) Yang-Mills QM of spatially constant fields}

The action of $SU(3$) Yang-Mills Quantum Mechanics of spatially constant gluon fields $V_{\mu}(t)\equiv V_{a\mu}(t) \lambda_a/2 $
is defined as
\begin{eqnarray}
{\cal S} [V] & : = & 
{\rm Vol} \int dt \left[ - \frac{1}{4} F^{\rm hom}_{a\ \mu\nu} F_a^{{\rm hom}\ \mu \nu}\right]
={\rm Vol} \int dt  \ \frac{1}{2}\left[ \left( E^{\rm hom}_{ai}\right)^2-  \left( B^{\rm hom}_{ai}\right)^2\right]~,
\label{action}
\end{eqnarray}
with the spatially constant field strength tensor
\begin{eqnarray}
F^{\rm hom}_{a\ \mu\nu} & : =& \delta_{0\mu} \partial_t V_{a \nu} 
 -  \delta_{0\nu} \partial_t V_{a \mu}+g f_{abc} V_{b \mu} V_{c \nu}~,\quad a=1,..,8~,
\end{eqnarray}
or in terms of the chromoelectric and chromomagnetic parts
\begin{eqnarray}
 E^{\rm hom}_{a i}\equiv F^{\rm hom}_{a\ i 0}\quad ,&&
\quad B^{\rm hom}_{a i} \equiv\frac{1}{2}\epsilon_{ijk}F^{\rm hom}_{a\ j k}~.
\end{eqnarray}
The action (\ref{action}) is invariant under the spatially homogeneous $SU(3)$ gauge transformations
\begin{eqnarray}
V_{a\mu}^{\omega}(t) \lambda_a/2  & =&
U[\omega(t)]\ \ V_{a\mu}(t) \lambda_a/2 \ \ U^{-1}[\omega(t)]~.
\end{eqnarray}
Furthermore, the action is invarinat under spatial rotations $R$
\begin{eqnarray}
 R:\quad   V_{a i}\rightarrow R_{ij} V_{a j}~,
\end{eqnarray}
as well as under parity transformations and charge conjugation 
\begin{eqnarray}
 P:\quad   V_{ai}\lambda_a  \rightarrow - V_{ai}\lambda_a  \quad\quad\quad
 C:\quad   V_{ai}\lambda_a  \rightarrow -( V_{ai}\lambda_a )^* ~.
\end{eqnarray}
In terms of the momenta
$\Pi_{ai}=-E_{ai}^{\rm hom}$ canonical conjugate to the spatial $V_{ai}$ one obtains the canonical Hamiltonian
\begin{eqnarray}
H_C&=&{\rm Vol}\ \Bigg[{1\over 2}\Pi_{ai}^2+{1\over 2} \left( B^{\rm hom}_{ai}(V)\right)^2
 -g  V_{a0} \left( f_{abc}V_{ci}  \Pi_{bi}\right)\Bigg]~.
\end{eqnarray}
Exploiting the  time dependence of the gauge transformations t put
\begin{equation}
V_{a0} = 0~,\quad\quad a=1,..,8 \quad\quad ({\rm Weyl\ gauge})~,
\end{equation}
the dynam. vaiables $V_{ai}$, $\Pi_{ai}$
are quantized in the Schr\"odinger functional approach imposing the equal-time commutation relations
 $\Pi_{ai} = -i\partial/\partial V_{ai}$.
The physical states $\Phi$ satisfy the coupled system of Schr\"odinger Equ. and eight non-abelian Gauss law constraints,
\begin{eqnarray}
  H_0\,\Phi &\equiv &
{\rm Vol}\  \left[{1\over 2} \Pi_{ai}^2+{1\over 2}\left(B^{\rm hom}_{ai}(V)\right)^2\right]\Phi=E\,\Phi~, 
\label{Sch-eq}\\
 G_a\,\Phi & \equiv&  g\ f_{abc}V_{ci}  \Pi_{bi}\ \Phi=0~, \quad\quad a=1,...8~.
\label{G-laws}
 \end{eqnarray}
The Gauss law operators $G_a$ are the generators of the residual  time independent gauge transformations,
satisfying $  [G_a,H]=0$ and $  [G_a,G_b]=if_{abc}G_c $.
The matrix element of an operator $O$ is given in the Cartesian form
\begin{eqnarray}
 \langle \Phi'| O|\Phi\rangle\ \propto \int dV\  \Phi'^*(V)\, O\, \Phi(V)~. 
 \end{eqnarray}
Since $H_0$ is invariant under spatial rotations $  [H_0,J_i]=0 $ with 
\begin{eqnarray}
J_i  =   \epsilon_{ijk}V_{aj} \Pi_{ak} \quad i=1,2,3~,\quad  [J_i,J_j]=i\epsilon_{ijk}J_k~,
\end{eqnarray}
and invariant under  parity  $ [H_0,P]=0$ 
and  charge conjugation $ [H_0,C]=0$ the eigenstates can be characterised by $ J^{PC}$.

In their work \cite{Weisz and Ziemann} Weisz and Ziemann used the variational approach to find the eigenvalues of the
constrained Schr\"odinger Equ. (\ref{Sch-eq}) with trial functions $$\Phi^{(J)PC}(V)=
P^{(J)PC}_{\rm gauge\ inv.}(V)\ \exp[-(\omega/2)\left(V_{ai}\right)^2]$$
which are gauge invariant and hence automatically satisfy the Gauss law constraints (\ref{G-laws}). 
In the sectors $0^{++}$ and $2^{++}$
they find rather accurate eigenvalues, in other sectors  first upper bounds.
In the present work we would like to demonstrate that the above constrained system becomes integrable if one replaces
\begin{eqnarray}
   \left(B^{\rm hom}_{ai}(V)\right)^2 
\ \longrightarrow \ \, \omega^2 \left(V_{ai}\right)^2~,\quad\quad  \omega>0\ \  {\rm  free\ parameter}
\nonumber
\end{eqnarray}
Using an exact gauge reduction the energy-eigensystem can be found rather accurately and used as a Hilbert-basis for the YM QM.
Truncating at higher and higher numbers of nodes, a converging low-energy  eigensystem of YM QM is obtained.

\section{Unconstrained Hamiltonian formulation using the flux-tube gauge}
\subsection{Unconstrained Hamiltonian formulation of SU(3) Yang-Mills QM}

In order to obtain an unconstrained Hamiltonian formulation, one can perform a point transformation of the original 24 $V_{ai}$  
to a new set of adapt coordinates, 
\begin{eqnarray}
V_{ai} \left(q, S \right) =
O_{ab}\left(q\right) A_{bi},
\end{eqnarray}
in terms of the  8 gauge angles $ q_j$ parametrising the $ O_{ab}(q)
\ {\rm orth.}\ 8\times 8\ {\rm matrix}\ {\rm adjoint \ to }\  U(q) $
\begin{eqnarray}
O_{ab}(q)=(1/8)\mbox{Tr}\left[U^{-1}(q)\lambda_a U(q)\lambda_b\right]~.
\end{eqnarray}
and 16 reduced $A_{ai}$ satisfying some gauge conditions
\begin{equation}
\chi_a(A)=\left(\Gamma_i\right)_{ab}\, A_{bi}=0~,\quad a=1,...,8~.
\nonumber
\end{equation}
Preserving the CCR $\rightarrow$ old canonical momenta in terms of the new variables
\begin{eqnarray}
 \Pi_{ai}(q,A,p,P)=O_{ab}\left(q\right)\left[ P_{bi}-\left(\Gamma_i\right)_{bl}
{\gamma}^{-1T}_{ls}(A)\left({1\over g}\Omega^{-1}_{st}(q) p_{t}
+T_s(A,P)\right)\right]~,
\end{eqnarray}
with the homogeneous part of the FP operator 
\begin{eqnarray}
\label{FP-op}
 \gamma_{ab}(A) := \left(\Gamma_i\right)_{ad} \, f_{dbc}A_{ci}~,
\end{eqnarray}
and the operators
\begin{eqnarray}
T_a(A,P) := f_{abc}A_{bi}P_{ci}~.
\label{Ta}
\end{eqnarray}
In terms of the new coordinates, the Gauss-laws become
\begin{eqnarray}
 G_a\Phi \equiv  O_{ak}(q)\Omega^{-1}_{ki}(q)  p_i\Phi
 =0\quad
\Leftrightarrow\quad\frac{\delta }{\delta q_i}\Phi=0\quad (\rm{Abelianisation})\nonumber
\end{eqnarray}
The unconstrained spin operator reads
\begin{eqnarray}
J_i=\epsilon_{ijk} A_{aj} E_{ak}~.
\label{Ji}
\end{eqnarray}
in terms of the  physical electric fields
\begin{eqnarray}
 E_{ai}:=P_{bi}-\left(\Gamma_i\right)_{bl}
{\gamma}^{-1T}_{ls}\! (A)\, T_s(A,P)~.
\label{Eai}
\end{eqnarray}

The correctly ordered unconstrained Hamiltonian of SU(3) YM-QM  takes the form  \cite{Christ and Lee}
\begin{eqnarray}
 H \!\!\!\!\! 
&=&\!\!\!\!\! {1\over 2}\Bigg[\frac{1}{|\gamma(A)|}\!\!
\ P_{ai}\ |\gamma(A)|\ P_{ai}+\frac{1}{|\gamma(A)|}\!\!
\ T_{a}\ |\gamma(A)|\left(\gamma^{-1}(\Gamma_i^T\Gamma_i)\gamma^{-1T}\right)_{ac} T_{c}
     +\left(B_{ai}^{\rm hom}(A)\right)^2\Bigg]~,
\nonumber
\end{eqnarray}
using the homogeneus part of the chromomagnetic field 
\begin{equation}
 B^{\rm hom}_{a\, i}(A):= (1/2) g\, \epsilon_{ijk}\, f_{abc}\, A_{b\, j}A_{c\, k}~.   
\end{equation}
The matrix element of a physical operator O is given by
\begin{eqnarray}
\langle \Psi'| O|\Psi\rangle\
\propto
\int dA\ 
 {|\gamma(A)|}\ \Psi'^*[A]\ O\ \Psi[A]~.
\nonumber
\end{eqnarray}

\subsection{Unconstrained Hamiltonian formulation of SU(3) YM-QM in the flux-tube gauge}

It is our aim to find a gauge which exists and leads to a maximally simple FP-operator.
This is can be acchieved by putting the "flux-tube-gauge", leading to a rather simple but non-trivial FP-operator.
The drawback of the this gauge is that the reduced fields $A_{ai}$ are color-singlets, but not spin-eigenstates,
as was the case for the SU(3) symmetric gauge \cite{pavel2012}- \cite{pavel2014} leading  to a very complicated Fp-operator.
We shall show that the disadvantage, that the reduced gauge fileds in the fluxtube-gauge are not spin eigenstates , 
can be circumvened by forming certain irreducible polynomials of the reduced $A$, symmetric tensors,
which have definite eigenvalues of J,P, and C.

The "flux-tube-gauge" is defined as
\begin{equation}
\chi_a(A)=0~:\quad\quad A_{a1}=0\quad \forall a=1,2,4,5,6,7 \quad \wedge  \quad A_{a2}=0\quad \forall a=5,7~.
\nonumber
\end{equation}
or explicitly
\begin{equation}
A=\,
 \left(
\begin{array}{c c c}
 0 & A_{12} & A_{13} \\ 
 0 & A_{22} & A_{23}\\ 
 A_{31} & A_{32} & A_{33} \\ 
 0 & A_{42} & A_{43} \\ 
 0 & 0 & A_{53}\\ 
 0 & A_{62} & A_{63} \\ 
 0 & 0 & A_{73} \\ 
 A_{81} & A_{82} & A_{83}    
\end{array}
\right)
\equiv  
\left( X\   Y\   Z  \right)~.  
\nonumber
\end{equation}
Using finally the reparametrisation
\begin{eqnarray}
 A_{31}\equiv  X_3 =r  \cos[\psi]  && A_{81}\equiv   X_8 =r \sin[\psi]~,
\end{eqnarray}
the explicit expression for the homogeneous part of the FP-operator (\ref{FP-op}) is
\begin{equation}
\gamma =
 \left(\!\!\!
\begin{array}{c c c c c c c c}
0  & -r  \cos[\psi]\!\!\!\! & 0 & 0 & 0 & 0 & 0 & 0 \\ 
r  \cos[\psi]\!\!\!\! & 0 & 0 & 0 & 0 & 0 & 0 & 0 \\ 
-Y_6/2 & 0 & -Y_4/2 &-Y_+  & 0 & Y_1/2 & -Y_2/2 &  - \sqrt{3}\, Y_4/2 \\ 
0 & 0 & 0 & 0 & r  \cos[\psi+{2\pi\over 3}]\!\!\!\! & 0 & 0 & 0 \\ 
0 & 0 & 0 &\!\!\!\! - r  \cos[\psi+{2\pi\over 3}] \!\!\!\! & 0 & 0 & 0 & 0 \\ 
0 & 0 & 0 & 0 & 0 & 0 &\!\!\!\! -  r  \cos[\psi+{4\pi\over 3}]\!\!\!\!  & 0 \\ 
0 & 0 & 0 & 0 & 0 &\!\!\!\!  r  \cos[\psi+{4\pi\over 3}]\!\!\!\!  & 0 & 0 \\ 
-Y_4/2 & 0 & Y_6/2 & Y_1/2 & Y_2/2 & Y_- & 0 & - \sqrt{3}\, Y_6/2  
\label{gamma}
\end{array}
\!\!\!\right)~,  
\end{equation}
using the abbreviations  $Y_\pm:=-(Y_3\pm \sqrt{3}Y_8)/2$. 
The  FP-determinant factorises
\begin{eqnarray}
\label{FP-det}
|\gamma(A)|=r^6 \cos^2[3\,\psi] \  Y_4 Y_6~.
\end{eqnarray}
The inverse $\gamma^{-1}$ of the Faddeev-Popov operator is rather simple (shown in Appendix A) and 
exists in the regions of non-vanishing
 determinant.  
The matrix elements are
\begin{equation}
\langle\Psi_1|O |\Psi_2\rangle =\int d\mu_X\int d\mu_Y\int d\mu_Z
  \ \Psi_1^{\dagger} O\ \Psi_2 ~,
\label{YM-measure}
\end{equation}
with the completely factorised
\begin{eqnarray}
\int d\mu_X &:= & \int_0^{\infty} d r\, r^{\, 7}  \int_0^{2\pi} d\psi\, \cos^2[3\,\psi]~, 
\label{muX}\\
\int d\mu_Y &:=&  
 \int_{-\infty}^{\infty} d Y_1\int_{-\infty}^{\infty} d Y_2\int_{-\infty}^{\infty} d Y_3\int_{-\infty}^{\infty} d Y_8
\int_{0}^{\infty} d Y_4 Y_4  \int_{0}^{\infty} d Y_6 Y_6~,
\label{muY}\\
\int d\mu_Z &:=& \prod_{a=1}^8\int_{-\infty}^{\infty} d Z_a~.
\label{muZ}
\end{eqnarray}

\noindent
Furthermore, for the operators $T_a(A,P)$ defined in (\ref{Ta}) we find that $T_a^{X}= 
 -i f_{abc}\, X_b\, \partial/\partial X_c  \equiv 0 ~$,
i.e.
\begin{equation}
T_a(A,P) \equiv T_a^{Y}(Y,P_Y)+T_a^{Z}(Z,P_Z)~.
\nonumber
\end{equation}
 where the components of the (non-reduced) $T_a^{Z} =\! -i f_{abc}\, Z_b\, \partial/\partial Z_c $ 
 satisfy the $su(3)$ algebra
\begin{equation}
[T^{Z}_a,T^{Z}_b]=i\, f_{abc}\, T^{Z}_c~,\nonumber
\end{equation}
whereas  the reduced $T_a^{Y} =\! -i f_{abc}\, Y_b\, \partial/\partial Y_c $ do not.

The physical electric fields read in the flux-tube gauge
\begin{equation}
E(A,P)=
 \left(
\begin{array}{c c c}
-{\cal P}_2(A,P) & P_{12} & P_{13} \\ 
{\cal P}_1(A,P)  & P_{22} & P_{23}\\ 
 P_{31} & P_{32} & P_{33} \\ 
{ \cal P}_5(A,P)  & P_{42} & P_{43} \\ 
-{ \cal P}_4(A,P)  & -{\cal P}_+(A,P) & P_{53}\\ 
- { \cal P}_7(A,P) & P_{62} & P_{63} \\ 
{ \cal P}_6(A,P) &  {\cal P}_-(A,P) & P_{73} \\ 
 P_{81} & P_{12} & P_{83}    
\end{array}
\right) ~, 
\end{equation}
with the Hermitean ${\cal P}_\pm^\dagger= {\cal P}_\pm$
\begin{eqnarray}
{\cal P}_+={1\over Y_4}\left[T^{Y}_3 +\left(T^{Z}_3+{1\over \sqrt{3}}T^{Z}_8 \right)\right]~,
& \quad\quad &
{\cal P}_-= {1\over Y_6}\left[T^{Y}_3 +\left(T^{Z}_3-{1\over \sqrt{3}}T^{Z}_8  \right)\right]~,
\end{eqnarray}
and
\begin{eqnarray}
 {\cal P}_1 \!\!\!\! & = & \!\!\!\! {1\over r\,\cos[\psi] }\left(\widetilde{T}^{Y}_1 +\widetilde{T}^{Z}_1\right)~,
\quad\quad\quad\quad\quad\quad\quad
 {\cal P}_2={1\over  r\,\cos[\psi]}\left(\widetilde{T}^{Y}_2 +\widetilde{T}^{Z}_2\right)~,
\nonumber\\
 {\cal P}_4\!\!\!\!&=&\!\!\!\! {1\over r\,\cos[\psi+2\pi/ 3] }\left(\widetilde{T}^{Y}_4 +\widetilde{T}^{Z}_4\right)~,
\quad\quad\quad
 {\cal P}_5= {1\over r\,\cos[\psi+2\pi/ 3] }\left(\widetilde{T}^{Y}_5 +\widetilde{T}^{Z}_5\right)~,
\nonumber\\
 {\cal P}_6\!\!\!\!&=&\!\!\!\! {1\over r\,\cos[\psi+4\pi/ 3] }\left(\widetilde{T}^{Y}_6 +\widetilde{T}^{Z}_6\right)~,
\quad\quad\quad
 {\cal P}_7= {1\over r\,\cos[\psi+4\pi/ 3] }\left(\widetilde{T}^{Y}_7 +\widetilde{T}^{Z}_7\right)~,
\end{eqnarray}
with the definition of the tilded operators $\widetilde{T}^{Y}$ and $\widetilde{T}^{Z}$ given in Appendix B. 

Hence, the components of the spin angular momentum operators  in reduced space 
\begin{eqnarray}
J_i=\epsilon_{ijk} A_{aj} E_{ak}~,
\label{Ji}
\end{eqnarray}
 read explicitly in the flux-tube gauge
\begin{eqnarray}
J_1& = &-i \!\!\!\!\sum_{a=1,2,3,4,6,8}\left[Y_a {\partial\over\partial Z_a}-Z_a {\partial\over\partial Y_a}\right]
                            +\left(Z_5{\cal P}_+-Z_7{\cal P}_-\right)~,
\nonumber\\
J_2& = &-i\sum_{a=3,8}\left[Z_a {\partial\over\partial X_a}-X_a {\partial\over\partial Z_a}\right]
                -\left(Z_1{\cal P}_2-Z_2{\cal P}_1\right)+\left(Z_4{\cal P}_5-Z_5{\cal P}_4\right)
               -\left(Z_6{\cal P}_7-Z_7{\cal P}_6\right)~,
\nonumber\\
J_3& = &-i\sum_{a=3,8}\left[X_a {\partial\over\partial Y_a}-Y_a {\partial\over\partial X_a}\right] 
              +\left(Y_1{\cal P}_2-Y_2{\cal P}_1\right)-Y_4{\cal P}_5 +Y_6{\cal P}_7~,
\label{J-flux-tube}
\end{eqnarray}
and the physical Hamiltonian reads
\begin{eqnarray}
H &=&{1\over 2}{\cal J}^{-1}_X\sum_{a=3,8}\left(
 {\partial\over  \partial X_{a}}{\cal J}_X{\partial\over  \partial X_{a}}\right)
+{1\over 2} {\cal J}_Y^{-1}  
\!\!\!\!\!\!\!\!
\sum_{a=1,2,3,4,6,8}\!\!\!
\left( {\partial\over  \partial Y_{a}}{\cal J}_Y{\partial\over  \partial Y_{a}}\right)
+ {1\over 2}\sum_{a=1}^8\left( {\partial\over  \partial Z_{a}}{\partial\over  \partial Z_{a}}\right)
\nonumber\\
&&\quad\quad
 + {1\over 2}{\cal J}_Y^{-1}
\!\!\!\!\!\!\!\!\!\!
\sum_{\alpha=1,2,4,5,6,7}
\!\!\!\!\!\!
{\cal P}^{\dagger}_{\alpha}\, {\cal J}_Y\, {\cal P}_{\alpha}
+ {1\over 2}\left({\cal P}_-^2+{\cal P}_+^2\right)
+ {1\over 2}\sum_{a=1}^8\sum_{i=1}^3  \left(B^{\rm hom}_{ai}[X,Y,Z]\right)^2~,
\label{H-flux-tube-a}
\end{eqnarray}
using the abbreviations $|\gamma(A)|={\cal J}_X\, {\cal J}_Y $ of (\ref{FP-det}) with 
${\cal J}_X:=r^6 \cos^2[3\,\psi]$ and ${\cal J}_Y:=Y_4 Y_6$.

\subsection{Hamiltonian of SU(3) YM-QM in the flux-tube gauge}

The Hamiltonian of SU(3) YM-QM in the flux-tube gauge (\ref{H-flux-tube-a}) can be written in the form 
\begin{eqnarray}
H[A,P]&=&K_X+K_Y+K_Z+   
{1\over 2\, r^2}\Bigg[\ {\left(I_1^{Y\!Z}+I_2^{Y\!Z}\right)\over  \cos^2{\psi}}
+\ {\left(I_4^{Y\!Z}+I_5^{Y\!Z}\right)\over  \cos^2{[\psi+2\pi/3]} }
+\ {\left(I_6^{Y\!Z}+I_7^{Y\!Z}\right)\over   \cos^2{[\psi+4\pi/3]} }
\Bigg]
\nonumber\\
&&
+{1\over 2\, Y_4^2 }I_+^{Y\!Z}
+{1\over 2\,  Y_6^2}I_-^{Y\!Z}\ 
+\ \  {1\over 2}\left(B^{\rm hom}_{ai}[X,Y,Z]\right)^2~.
\label{H-flux-tube}
\end{eqnarray}
The single-direction kinetic terms read
\begin{eqnarray}
K_X &= &
-{1 \over 2 }\left[{\partial^2\over\partial r^2}+ { 7\over r} {\partial\over\partial r}+
 {1 \over r^2  } \left(-6 \tan[3 \psi]{\partial\over\partial \psi}+  {\partial^2\over\partial\psi^2}\right)\right]~,
\nonumber\\
K_Y  &=&
-{1 \over 2 }\Bigg[\sum_{a=1,2,3,8} {\partial^2\over  \partial Y_{a}^2}
+\sum_{a=4,6} \left({\partial^2\over\partial Y_a^2} +{ 1\over Y_a } {\partial\over\partial Y_a}
-{1\over   Y_a^2} \left(Y_1  {\partial\over\partial Y_2}-Y_2  {\partial\over\partial Y_1}\right)^2 \right)
\Bigg]~,
\nonumber
\\
 K_Z & =& -{1 \over 2 }\sum_{a=1}^8 {\partial^2\over  \partial Z_{a}^2} ,
\nonumber
\end{eqnarray} 
and the interations $I_m^{Y\!Z}$ and $I_{\pm}^{Y\!Z}$ given in Appendix B.

It has been proven a long time ago by Simon \cite{Simon}, that the spectrum is discrete although the chromomagnetic potential
owns three flat valleys narrowing down. Although there exist classical zero energy trajectories to infinity,
 the quantum fluctuations in the
narrowing valleys confine the wavefunction and lead to a discret spectrum. It is therefore reasonable to replace in an intermediate
step the chromomagnetic potential by a separable 16-dimensional harmonic oscillator potential, and then use the obtained
eigensystem to find the eigensystem of SU(3) YM QM.

\subsection{The corresponding harmonic oscillator problem $H_{h.o.}$}

Replacing in $ H (A,P)$ the magnetic potential by the separable harmonic oscillator potential
 with free parameter $\omega>0$
\begin{eqnarray}
 {1\over 2}   \left(B^{\rm hom}_{ai}(A)\right)^2 
\ \longrightarrow \  {1\over 2}\, \omega^2 \left(A_{ai}\right)^2
\equiv   {1\over 2}\, \omega^2\left[r^2+Y_1^2+Y_2^2+Y_3^2+Y_4^2+Y_6^2+Y_8^2+Z_a^2\right]
\label{B2toA2}
\end{eqnarray}
we obtain the corresponding harmonic oscillator problem (with the same measure !!!),
\begin{eqnarray}
H_{h.o.}[A,P]&=&H_X+H_Y+H_Z+ {1\over 2\, Y_4^2 }I_+^{Y\!Z}
+{1\over 2\,  Y_6^2}I_-^{Y\!Z}+\ 
  \nonumber\\
&&
+{1\over 2\, r^2}\Bigg[\ {\left(I_1^{Y\!Z}+I_2^{Y\!Z}\right)\over  \cos^2{\psi}}
+\ {\left(I_4^{Y\!Z}+I_5^{Y\!Z}\right)\over  \cos^2{[\psi+2\pi/3]} }
+\ {\left(I_6^{Y\!Z}+I_7^{Y\!Z}\right)\over   \cos^2{[\psi+4\pi/3]} }
\Bigg]
\label{H_{h.o.}}
\end{eqnarray}
The single-direction Hamiltonions read
\begin{eqnarray}
H_X  \!\!\!\!\!\!&= &\!\!\!\!\!\!
{1 \over 2 }\left[-{\partial^2\over\partial r^2}- { 7\over r} {\partial\over\partial r}+
 {1 \over r^2  } \left(6 \tan[3 \psi]{\partial\over\partial \psi}-  {\partial^2\over\partial\psi^2}\right)+\omega^2 r^2\right]~,
\nonumber\\
H_Y \!\!\!\!\!\! &=& \!\!\!\!\!\!
{1 \over 2 }\Bigg[\sum_{a=1,2,3,8}\left( -{\partial^2\over  \partial Y_{a}^2}+\omega^2 Y_{a}^2\right)
+\sum_{a=4,6} \left(-{\partial^2\over\partial Y_a^2} -{ 1\over Y_a } {\partial\over\partial Y_a}
+{1\over   Y_a^2} \left(Y_1  {\partial\over\partial Y_2}-Y_2  {\partial\over\partial Y_1}\right)^2+\omega^2 Y_{a}^2 \right)
\Bigg]~,
\nonumber\\
 H_Z  \!\!\!\!\!\!& =& \!\!\!\!\!\! {1 \over 2 }\sum_{a=1}^8 \left[-{\partial^2\over  \partial Z_{a}^2} +\omega^2 Z_{a}^2\right] ,
\nonumber
\end{eqnarray} 
As stated already in our earlier work \cite{pavel2016}, this system is integrable and can be solved analytically in terms of
orthogonal polynomials. We shall demonstrate this in more detail in the following pragraphs.

\section{ Exact solution of the corresponding harmonic oscillator problem}

The operators  $T^Y_a$ and $T^Z_a$ lead to the coupling between the three spatial directions.

\subsection{Solutions of the $H_{h.o.}$ Schr\"odinger equation separable in $X,Y,Z$.}

First looking for solutions for the case where the kinetic terms decouple for all three directions:
\begin{eqnarray}
&& \Phi_{X|Y|Z}= \Phi_{X}[X]\Phi_{Y}[Y]  \Phi_{Z}[Z]~, 
\nonumber\\
 H_{h.o.}\Phi_{X|Y|Z}&=&\left(H_X\! +\! H_Y\! +\! H_Z \right)\Phi_{X|Y|Z}\ = \left( \epsilon_X+ \epsilon_Y+ \epsilon_Z\right) 
 \Phi_{X|Y|Z}~. 
\end{eqnarray}
with the single-direction functionals $ \Phi_{X}$, $\Phi_{Y}$, and $ \Phi_{Z}$  satisfying the separate Schr\"odinger equations
\begin{eqnarray}
H_X \Phi_X &=& \epsilon_X \Phi_X~,
\label{X-equation}\\   
H_Y \Phi_Y &=& \epsilon_Y \Phi_Y  \quad \wedge \quad  T^Y_a \Phi_Y  =0~, \quad a=1,...,8~,
\label{Y-equation}
\\   
H_Z \Phi_Z &=& \epsilon_Z \Phi_Z  \quad \wedge \quad  T^Z_a \Phi_Z  =0~,   \quad a=1,...,8~.
\label{Z-equation}
\end{eqnarray}
Note that the X-equation is unconstrained, whereas the Y- and Z-equations are constrained.

\subsubsection{Solution of the X-equation}

Consider first the X-equation (\ref{X-equation}) 
\begin{equation}
H_X \Phi_X \equiv
 -{1 \over 2 }\left[{\partial^2\over\partial r^2}+ { 7\over r} {\partial\over\partial r}
- \omega^2 r^2+ {1 \over r^2  } \left(-6 \tan[3\, \psi]{\partial\over\partial \psi}+  {\partial^2\over\partial\psi^2}\right)
\right] \Phi_X =
 \epsilon_X \Phi_X
\label{X-Equ}
\end{equation}
and the matrix elements
\begin{equation}
\label{X-measure}
\langle \Phi^\prime_X |O_X|\Phi_X\rangle= 
 \int_0^{\infty}\!\!\!\! dr\,  r^7\int_0^{2\pi}\!\!\!\! d\psi\, \cos^2[3\,\psi] \ \ \Phi^\prime_X \ O_X[X]\ \Phi_X
\end{equation}
In terms of the new coordinates
\begin{equation}
\label{s11,s111}
  s_{11}=X_{a}X_{a}= r^2~,\quad\quad\quad\quad s_{111}=d_{abc} X_{a}X_{b}X_{c}
 ={1\over \sqrt{3}} r^3 \sin[3 \psi]    
\end{equation}
which are the $s_{11}$ and  $s_{111}$ components of the 6- and 10-component symmetric tensors
\begin{equation}
 s^{++}_{[2]ij}[A]:=A_{a i}A_{a j}~,\quad\quad\quad\quad s^{--}_{[3]ijk}[A]:=d_{abc} A_{a i}A_{b j}A_{c k}
\label{s_ij,s_ijk}
\end{equation}
and using the scaled 
\begin{equation}
     \overline{y}:=s_{111}/s_{11}^{3/2}= {1\over \sqrt{3}}\sin[3\,\psi]    
\nonumber
\end{equation}
Equ. (\ref{X-Equ}) reads
\begin{equation}
-{1 \over 2 }\left[4 s_{11} {\partial^2\over\partial s_{11}^2}+16 {\partial\over\partial s_{11}} - s_{11}\ \omega^2+
{1 \over s_{11}}\left(3 (1 -3  \overline{y}^2) {\partial^2\over\partial  \overline{y}^2} 
- 27 \overline{ y} {\partial\over\partial  \overline{y}}\right) 
\right] \Phi_X =
 \epsilon_X \Phi_X
\nonumber
\end{equation}
which can easily be solved by separation of variables.
The solutions can be written 
\begin{equation}
\Phi^X_{n_1,n_2}[s_{11},s_{111} ] = 
 {\omega^2 \over
 \sqrt{6\pi}}\  p^{(1)}_{n_1,n_2}\left(\omega\, s_{11},\omega^{3/2}\, s_{111}\right)  \exp{[-\omega\, s_{11}/2 ]}
\label{X-equ-sol}
\end{equation}
 with the energy eigenvalues
\begin{equation}
   \epsilon_{n_1,n_2}=\left(4+2 n_1+3 n_2\right)\omega =:\left(4+n_X\right)\omega   
\end{equation}
The lowest polynomials read
\begin{eqnarray}
&& p^{(1)}_{0,0}(x,y)=1~,\quad\quad  p^{(1)}_{1,0}(x,y)= { 1 \over 2}(-4 + x)~,
\quad\quad   p^{(1)}_{0,1}(x,y)= { 1 \over \sqrt{10}}\, y ~, \\
\nonumber
&& p^{(1)}_{2,0}(x,y)= { 1 \over 2 \sqrt{10}}(20-10 x + x^2)~,
\quad\quad  p^{(1)}_{1,1}(x,y)= { 1 \over  \sqrt{70}}(-7 + x) y~,\quad ... 
\end{eqnarray}
The number $n_X:=2 n_1+3 n_2$ is the degree of the corresponding polynomial in the components of $X$ for each solution.
After Gram-Schmidt-orthogonalisation of those degenerate in energy, they an ONB with respect to the measure (\ref{X-measure}).

\noindent
We mention here, that the X-equation (\ref{X-equation}), due to the special form of the FP-Jacobian (\ref{FP-op}),
can be considered also for the case of singular boundary conditions. The lowest solutions for this singular case and their
energy eigenvalues are shown in Appendix C.
Although very interesting, we shall limit ourselves in this article on the regular case.

\subsubsection{Solution of the Y-equation}

For functonals $ \Phi_Y[s_{22},s_{222}]$ depending only on the components
\begin{eqnarray}
s_{22}=Y_a Y_a~,\quad\quad\quad\quad s_{222}= d_{abc} Y_{a }Y_{b }Y_{c } ~, 
\end{eqnarray}
of the symmetric tensors (\ref{s_ij,s_ijk}), satisfying $T^Y_a  s_{22} =
 T^Y_a  s_{222} =0$, i.e $ T^Y_a\Phi_Y[s_{22},s_{222}]=0~, a=1,...,8,$ and the Y-Equ.(\ref{Y-equation})
\begin{eqnarray}
 H_Y \Phi_Y[s_{22},s_{222}] = \epsilon_Y \Phi_Y[s_{22},s_{222}]
\end{eqnarray}
is solved by the same functionals (\ref{X-equ-sol}), and in particular polynomials, and energy eigenvalues 
as for the X-Equ.(\ref{X-equation}), but with $s_{11}$ and  $s_{111}$ replaced
by $s_{22}$ and  $s_{222}$ respectively.
They form an ONB with respect to the measure
\begin{equation}
\langle \Phi^\prime_Y |O_Y|\Phi_Y\rangle= 
 \int_0^{\infty} dY_4 Y_4  \int_0^{\infty} dY_6 Y_6 \prod_{a=1,2,3,8}\left[ \int_0^{\infty} dY_a \right]
 \  \Phi^\prime_Y \, O_Y[Y]\, \Phi_Y~.
\end{equation}

\subsubsection{Solution of the Z-equation}

Similarly for the Z-equation with $ \Phi_Z[s_{33},s_{333}]$ depending only on the components 
\begin{eqnarray}
s_{33}=Z_a Z_a~,\quad\quad\quad\quad s_{333} = d_{abc} Z_{a }Z_{b }Z_{c } ~, 
\end{eqnarray}
of the symmetric tensors (\ref{s_ij,s_ijk}), satisfying $T^Z_a  s_{33} =
 T^Z_a  s_{333} =0$, i.e.  $T^Z_a\Phi_Z[s_{33},s_{333}]=0~, a=1,...,8~,$ and the Z-Equ.(\ref{Z-equation})
\begin{eqnarray}
 H_Z \Phi_Z[s_{33},s_{333}] = \epsilon_Z\Phi_Z[s_{33},s_{333}]~,
\end{eqnarray}
solved by the same functionals (\ref{X-equ-sol}), and in particular polynomials, and energy eigenvalues 
as for the X-Equ.(\ref{X-equation}), but with $s_{11}$ and  $s_{111}$ replaced
by $s_{33}$ and  $s_{333}$ respectively.
They form an ONB with respect to the measure
\begin{equation}
\langle \Phi^\prime_Z |O_Z|\Phi_Z\rangle= 
 \prod_{a=1,...,8}\left[ \int_0^{\infty} dZ_a \right]
 \  \Phi^\prime_Z \, O_Z[Z]\, \Phi_Z~.
\end{equation}

\subsubsection{Trigonal form of the one-direction Hamiltonian}
The polynomials can be easily obtained using the equation
\begin{eqnarray}
D^{(1)}\  p^{(1)}_{n_1, n_2 }(x,y)  =(\epsilon/\omega)\  p^{(1)}_{n_1, n_2 }(x,y)~, 
\end{eqnarray}
with the differential operator $D^{(1)} := D_{-2}^{(1)}+D_{0}^{(1)}$ consisting of two parts
\begin{eqnarray}
D_{-2}^{(1)}:= -{3\over 2}x^2\partial_y^2-2\left(x\partial_x+3y\partial_y+4\right)\partial_x~,&\quad &
D_{0}^{(1)}:=\left(2x\partial_x+3y\partial_y+4\right)~, 
\label{D1}
\end{eqnarray}
acting in the space of monomials $x^{n_1} y^{n_2}$. The operator $D_{0}^{(1)}$ reproduces the given
monomial with eigenvalue $(2n_1+3 n_2+4)$ where $(2n_1+3 n_2)$ is its power seen as a 
homogeneous polynomial in the $A$ ("A-power"),
attributing $x$ the A-power $2$ and $y$ the A-power $3$, as is the case e.g. for $x=\omega s_{11}(A)$ and 
 $y=\omega^{3/2} s_{111}(A)$ in Equ.(\ref{s11,s111}).
On the other hand, the operator $D_{=2}^{(1)}$ transforms one $x^{n_1} y^{n_2}$ into another one, $x^{m_1} y^{m_2}$,
with an by two lowered A-power $(2m_1+3 m_2)=(2n_1+3 n_2)-2$. 
Hence the operator $D^{(1)}$ is triangular in the space of monomials $x^{n_1} y^{n_2}$ and 
can easily be diagonalised with eigenvalues
\begin{eqnarray}
\epsilon=(2n_1+3 n_2+4)\omega~,
\end{eqnarray}
and the eigenfunctions the polynomials
\begin{eqnarray}
p^{(1)}_{n_1, n_2}(x,y)={\cal N}\left(
\sum_{m_1,m_2}^{2 m_1+3 m_2 < 2 n_1+3 n_2}
\!\!\!\!\!\!\!\!\!\!\!\! \left[ {\cal A}^{(1)}_{n_1,n_2 }(m_1,m_2)\ x^{m_1} y^{m_2}\right]+x^{n_1} y^{n_2}\right)~,
\end{eqnarray}
with the leading ("defining") monomial $x^{n_1} y^{n_2}$  and a "tail" of monomials of decreasing, lower powers 
with some definite coefficients ${\cal A}^{(1)}_{n_1,n_2 }(m_1,m_2)$.

We shall see in the folling paragraphs, that such behaviour will also for the case of solutions separable in only one space
direction and the non separable general case. Before discussing these we shall give in the following subsection some
examples of solution build from those separable in all $X,Y,Z$.

\subsubsection{Putting the solutions of the X-,Y, and Z-equations together}

Together, we have
 \begin{equation}
 H_{h.o.}\Phi_{X|Y|Z}= \left( \epsilon_X+ \epsilon_Y+ \epsilon_Z\right) =\epsilon_{h.o.}\, \Phi_{X|Y|Z}~. 
\end{equation}
with the energy eigenvalues
\begin{eqnarray}
\epsilon_{h.o.}=\left(12+n_X+n_Y+n_Z\right)\omega=\left(12+n\right)\omega
\end{eqnarray}
We find the lowest solutions
$
\ \Phi_{[n] X|Y|Z}[A]= P_{[n] X|Y|Z} \exp[-\omega\left(s_{11}+s_{22}+s_{33}\right)/2]
$
with
\begin{eqnarray}
\epsilon_{h.o.} = 12\,\omega~: &&\!\!\!\!\!\!
P_{[0]\, X|Y|Z} \propto  p_{0, 0}[X]\, p_{0, 0}[Y]\, p_{0, 0}[Z]= 1~, 
\nonumber
\\ 
\epsilon_{h.o.} = 14\,\omega~: &&\!\!\!\!\!\!
P_{[2]\, X|Y|Z}\propto p_{1, 0}[X]\, p_{0, 0}[Y]\, p_{0, 0}[Z]= \left(-2+\omega s_{11}/2\right)~,\quad {\rm and\ perm.}  
\nonumber
\\ 
\epsilon_{h.o.} = 15\,\omega~: &&\!\!\!\!\!\!
P_{[3]\, X|Y|Z}\propto p_{0, 1}[X]\, p_{0, 0}[Y]\, p_{0, 0}[Z]= \omega^{3/2} s_{111}/\sqrt{10}~,\quad {\rm and\ perm.} 
\nonumber
\\ 
\epsilon_{h.o.} = 16\,\omega~:&&\!\!\!\!\!\!
 P_{[4]1\, X|Y|Z} \propto p_{2, 0}[X]\, p_{0, 0}[Y]\, p_{0, 0}[Z]
= \left(10-5\,\omega s_{11}+\omega^2 s_{11}^2/2\right)/\sqrt{10}~,
\ {\rm and\ perm.} 
\nonumber
\\ 
&&\!\!\!\!\!\! P_{[4]2\, X|Y|Z}\propto p_{0, 0}[X]\, p_{1, 0}[Y]\, p_{1, 0}[Z]
=  4-\omega (s_{22}+s_{33})+\omega^2 s_{22} s_{33}/4~,
\ {\rm and\ perm.}
\nonumber
\\ 
\epsilon_{h.o.} = 17\,\omega~:&&\!\!\!\!\!\!
 P_{[5]1\, X|Y|Z} \propto p_{1, 1}[X]\, p_{0, 0}[Y]\, p_{0, 0}[Z]
= \left(-7+\omega s_{11}\right)\omega^{3/2}\, s_{111}/\sqrt{70}~,
\ {\rm and\ perm.} 
\nonumber
\\ 
&&\!\!\!\!\!\! P_{[5]2\, X|Y|Z}\propto p_{0, 1}[X]\, p_{1, 0}[Y]\, p_{0, 0}[Z]=
 \left( -2+\omega s_{22}/2\right)\omega^{3/2} s_{111}/\sqrt{10} ~,
\ {\rm and\ perm.}
\nonumber
\\ 
...
\label{X|Y|Z-sol}
\end{eqnarray}
The number $n$ is the maximal power of the polynomial $P_[n]$, seen as a polynomial in the reduced gauge field $A$.
Superposing (\ref{X|Y|Z-sol}), and denoting 
$$
 s_{[2]}^{(0)}:=s_{[2]ii}=s_{11} +s_{22}+s_{33}=  \left(A_{ai}\right)^2
$$
we can build the $0^{++}$ eigenstates
$
 \Phi_{[n]}^{(0)++}[A]= P_{[n]}^{(0)++}[A] \exp[-\omega\left(A_{ai}\right)^2/2],
$   
with
\begin{eqnarray}
\epsilon^{(0)++}_{h.o.} = 12\,\omega~:
\quad 
P_{[0]}^{(0)++}\!\!\! &\propto&\!\!\! 1~, 
\nonumber\\
\epsilon^{(0)++}_{h.o.} = 14\,\omega~:
\quad 
P_{[2]}^{(0)++}\!\!\! &\propto&\!\!\!  -12+\omega\, s^{(0)}_{[2]}~,    
\nonumber\\
\epsilon^{(0)++}_{h.o.} = 16\,\omega~:
\quad 
P_{[4]}^{(0)++}\!\!\! &\propto&\!\!\!  108 -18\,\omega\,  s^{(0)}_{[2]}
+\,\omega^2  \!  \left(s^{(0)}_{[2]}\right)^2 .
\label{Spin-0-sol}
\end{eqnarray}
We find rotational invariant solutions although the flux-tube gauge is not rotational invariant.

\noindent
We mention here, that for the case of singular solutions of the the X-equation (\ref{X-equation}), 
discussed in Appendix C, rotational invariance 
is broken in one direction, here the x-direction, leaving only a cylindrical symmetry.

\subsection{Separation of one of the three directions }

\subsubsection{Separation of the Y-equation}

Starting with the case where the y-direction decouples
\begin{eqnarray}
\Phi_{XZ|Y}[X,Y,Z]=  \Phi_{Y}[Y] \   \Phi_{XZ}[X,Z]~,   
\nonumber 
\end{eqnarray}
\vspace{-0.9cm}
\begin{eqnarray}
H_{h.o.}\Phi_{Y|XZ}[X,Y,Z]=\left(H_Y+H_{XZ}\right)\Phi_{Y|XZ}[X,Y,Z]=(\epsilon_Y+\epsilon_{XZ})\Phi_{Y|XZ}[X,Y,Z]~.
\end{eqnarray}
where $ \Phi_{Y} $ satisfies the above discussed constrained Y-equation (\ref{Y-equation})
and $ \Phi_{XZ}$  is to solve the constrained x-z-direction Schroedinger equation 
\begin{eqnarray}
 \left[H_X+H_Z+{1\over 2\, r^2}\!\!\left[
{\left(T_1^Z\right)^2+\left(T_2^Z\right)^2\over  \cos^2{\psi}}
+{\left(T_4^Z\right)^2+\left(T_5^Z\right)^2\over  \cos^2{[\psi+2\pi/3]}}
+{\left(T_6^Z\right)^2+\left(T_7^Z\right)^2\over  \cos^2{[\psi+4\pi/3]}}
\right]\!\right]\!\! \Phi_{XZ} &=&
 \epsilon_{XZ} \Phi_{XZ}~,
\nonumber \\
   \wedge  \quad   T_3^{Z} \Phi_{XZ}=0 \quad  \wedge\quad    T_8^{Z} \Phi_{XZ}= 0~.
\quad\quad\quad \quad\quad\quad\quad \quad &&
\end{eqnarray}

\noindent
In terms of the components $s_{13}$, $s_{113}$, and  $s_{133}$ of the symmetric tensor  (\ref{s_ij,s_ijk}),
\begin{eqnarray}
s_{13}=X_{a} Z_{a}~,
\quad\quad s_{113}=d_{abc } X_{a} X_{b} Z_{c}~,
\quad\quad
s_{133}= d_{abc } X_{a} Z_{b} Z_{c} ~,
\end{eqnarray}
we find the $\epsilon_{XZ}=10\,\omega$ solution
\begin{eqnarray}
\Phi^{XZ}_{[2]}[X,Z] & \propto & 
 \big[ \omega\,  s_{13}\big] \exp{[-\omega (s_{11}+s_{33})/2 ]}~,
\end{eqnarray}
and the $\epsilon_{XZ}=11\,\omega$ solutions
\begin{eqnarray}
\Phi^{XZ}_{[3]1}[X,Z] &\propto &  
\Big[ \omega^{3/2}\,  s_{113}\Big] \exp{[-\omega (s_{11}+s_{33})/2 ]}~,
\nonumber\\
\Phi^{XZ}_{[3]2}[X,Z] & \propto &
 \Big[ \omega^{3/2}\,  s_{133}\Big] \exp{[-\omega (s_{11}+s_{33})/2 ]}~,
\end{eqnarray}

\noindent
We have the $\epsilon^{(2)}=12\,\omega$ solutions
\begin{eqnarray}
\Phi^{XZ}_{[4]1}[X,Z] & \propto & 
\Big[2- {1\over 2} \omega ( s_{11}+ s_{33})+ \omega^2  s_{13}^2 \Big] \exp{[-\omega (s_{11}+s_{33})/2 ]}~,
\\
\Phi^{XZ}_{[4]2}[X,Z] & \propto & 
 \Big[6-{3\over 2}\omega ( s_{11}+ s_{33}) + \omega^2  b_{22} \Big] \exp{[-\omega (s_{11}+s_{33})/2 ]}~,
\end{eqnarray}
using additionally the $22$-component 
\begin{eqnarray}
 b_{22}[X,Z] =f_{abc}\, f_{ade}\, X_b\, Z_c\, X_d\, Z_e~,
\end{eqnarray}
 of the symmetric tensor
\begin{eqnarray}
 b^{++}_{[4]ij}[A]& :=&g^{-2} B^{\rm hom}_{a\, i}[A]\, B^{\rm hom}_{a\, j}[A] ~,
\label{b_ij}
\end{eqnarray}
It appears in the tail of the  $\epsilon^{(2)}=14\,\omega$ solution  
\begin{eqnarray}
\Phi^{XZ}_{[6]}[X,Z] & \propto &
  \Big[
\omega^2\left({1\over 2} s_{11}s_{33}-s_{13}^2- b_{22}\right)
+ \omega^3 s_{111} s_{133} 
  \Big] \exp{[-\omega (s_{11}+s_{33})/2 ]}~,
\end{eqnarray}
with the product $s_{111} s_{133}$ as leading term, and is irreducible in A-space.

In general, the X-Z-solutions are of the form
\begin{eqnarray}
\!\!\!\!\!\!\!\!\!\!\!\!\!\!\!
\Phi^{XZ}_{n_1,...,n_8}[X,Z] \!\!\!\!\! & =& \!\!\!\!\!
  {2\over 5} {\omega^4 \over \sqrt{11} \pi^{  3/2}}
 p^{(2)}_{n_1,...,n_8}\left(\omega s_{11},\omega s_{33},\omega s_{13},\omega^{3/2} s_{111},\omega^{3/2} s_{333},
\omega^{3/2} s_{113},\omega^{3/2} s_{133},\omega^{2}b_{22}\right)
\nonumber\\
&&\quad\quad\quad\quad\quad\quad
\times\exp{[-\omega (s_{11}+s_{33})/2 ]}
\end{eqnarray}
with energy
\begin{eqnarray}
\epsilon^{(2)}_{n_1,...,n_8}:=\left[8+2(n_1+n_2)+3(n_3+n_4+n_5+n_6+n_7)+4n_8\right]\omega
\end{eqnarray} 
where $ p^{(2)}_{n_1,...,n_8}\left(x_1,...,x_8\right)$ is a polynomial in the eight variables $x_1,...,x_8$ where the monomial 
with the maximal order reads $\prod_{i=1}^8 x_i^{n_i}$.
After orthogonalisation of those degenerate in energy) they form an ONB with respect to the measure
\begin{equation}
\langle \Phi^\prime_{XZ} |O|\Phi_{XZ} \rangle= 
\int d\mu_X \int d\mu_Z
\ \  \Phi^\prime_{XZ}  \, O[X,Z]\, \Phi_{XZ} ~.
\end{equation}

\subsubsection{Separation of the Z-equation}

Next we consider the case where the z-direction decouples
\begin{eqnarray}
\Phi_{Z|XY}[X,Y,Z] =  \Phi_{Z}[Z] \   \Phi_{XY}[X,Y]~,   
\nonumber 
\end{eqnarray}
\vspace{-0.7cm}
\begin{eqnarray}
H_{h.o.}\Phi_{Z|XY}[X,Y,Z] =\left(H_Z+H_{XY}\right)\Phi_{Z|XY}[X,Y,Z]=(\epsilon_Z+\epsilon_{XY})\Phi_{Z|XY}[X,Y,Z]~.
\end{eqnarray}
where $ \Phi_{Z} $ satisfies the above discussed constrained Z-Schr\"odinger-equation (\ref{Z-equation})
and functional $ \Phi_{XY}$ is to solve the unconstrained x-y-direction Schr\"odinger equation
\begin{eqnarray}
\!\!\!\!\!\!\!\!\!\!\!\!\!\!
H_{XY}\Phi_{XY} \!\!\!\! &\equiv &\!\!\!\!   \Bigg[H_X+H_Y+{1\over 2\, r^2 ( Y_4 Y_6)}\Bigg[
{\widetilde{T}_1^{Y\dagger}\, Y_4 Y_6\,  \widetilde{T}^{Y}_1 
+\widetilde{T}_2^{Y\dagger}\, Y_4 Y_6 \,  \widetilde{T}^{Y}_2\over  \cos^2{\psi}}+
\nonumber\\
&&
+{\widetilde{T}_4^{Y\dagger}\,  Y_4 Y_6\,  \widetilde{T}^{Y}_4 
+\widetilde{T}_5^{Y\dagger}\, Y_4 Y_6\,  \widetilde{T}^{Y}_5\over  \cos^2{[\psi+2\pi/3]}}
+{\widetilde{T}_6^{Y\dagger}\, Y_4 Y_6\,  \widetilde{T}^{Y}_6] 
+\widetilde{T}_7^{Y\dagger}\,  Y_4 Y_6\,  \widetilde{T}^{Y}_7\over  \cos^2{[\psi+4\pi/3]}}
\Bigg]\Bigg] \Phi_{XY}
= \epsilon_{XY} \Phi_{XY}
\end{eqnarray}
Using also the additional components
\begin{eqnarray}
 s_{12}= X_{a } Y_{a }~,\quad
s_{112}=d_{abc } X_{a } X_{b } Y_{c }~, \quad
s_{122}=d_{abc } X_{a } Y_{b } Y_{c }~,\quad
 b_{33}[X,Y]=f_{abc}\, f_{ade}\, X_b\, Y_c\, X_d\, Y_e~, 
\end{eqnarray}
the x-y-solutions are of the form
\begin{eqnarray}
\!\!\!\!\!\!\!\!\!\!\!\!\!\!\!
\Phi^{XY}_{n_1,...,n_8}[X,Y] \!\!\!\!\! & =& \!\!\!\!\!
  {2\over 5} {\omega^4 \over \sqrt{11} \pi^{  3/2}}
 p^{(2)}_{n_1,...,n_8}\left(\omega s_{11},\omega s_{22},\omega s_{12},\omega^{3/2} s_{111},\omega^{3/2} s_{222},
\omega^{3/2} s_{112},\omega^{3/2} s_{122},\omega^{2}b_{33}\right)
\nonumber\\
&&\quad\quad\quad\quad\quad\quad
\times\exp{[-\omega (s_{11}+s_{22})/2 ]}
\end{eqnarray}
and energy
\begin{eqnarray}
\epsilon^{(2)}_{n_1,...,n_8}:=\left[8+2(n_1+n_2)+3(n_3+n_4+n_5+n_6+n_7)+4n_8\right]\omega
\end{eqnarray} 
where $ p^{(2)}_{n_1,...,n_8}$ are the same polynomials as for the x-z case.
After orthogonalisation of those degenerate in energy) they form an ONB with respect to the measure
\begin{equation}
\langle \Phi^\prime_{XY} |O|\Phi_{XY}\rangle=
\int d\mu_X \int d\mu_Y
\ \  \Phi^\prime_{XY}  \, O[X,Y]\, \Phi_{XY} ~.
\end{equation}

\subsubsection{Separation of the X-equation}
Finally,  we consider the case where the x-direction decouples
\begin{eqnarray}
\Phi_{X|YZ}[X,Y,Z]=  \Phi_{X}[X] \   \Phi_{YZ}[Y,Z]~,   
\nonumber 
\end{eqnarray}
\vspace{-0.7cm}
\begin{eqnarray}
H_{h.o.}\Phi_{X|YZ}[X,Y,Z]=\left(H_X+H_{YZ}\right)\Phi_{X|YZ}[X,Y,Z]=\left(\epsilon_X+\epsilon_{YZ}\right)\Phi_{X|YZ}[X,Y,Z]~.
\end{eqnarray}
where $ \Phi_{X} $ satisfies the above discussed unconstrained X-equation (\ref{X-equation})
and the functionals $ \Phi_{YZ}$ have to solve the constrained  Y-Z- equation
\begin{eqnarray}
H_{YZ} \Phi_{YZ}\equiv  \Bigg[H_Y+H_Z
  \!\!\!\!   &+&\!\!\!\!      {1\over 2 Y_4^2}\left[
            {1\over 2}\left( T^Z_3+{1\over \sqrt{3}} T^Z_8 \right)- T^Y_3 
 \right]
\left( T^Z_3+{1\over \sqrt{3}} T^Z_8\right)
\nonumber\\
&&
+{1\over 2 Y_6^2}\left[ {1\over 2}\left( T^Z_3-{1\over \sqrt{3}} T^Z_8 \right)- T^Y_3 
 \right]
\left( T^Z_3-{1\over \sqrt{3}} T^Z_8\right)
           \Bigg] \Phi_{YZ}= \epsilon_{YZ}\, \Phi_{YZ}~,
\nonumber\\
\wedge\quad  \left( \widetilde{T}^Y_a + \widetilde{T}^{Z}_a\right) \Phi_{YZ} &=& 0 ~, \quad a=1,2,4,5,6,7~.
\end{eqnarray}
In terms of the components
\begin{eqnarray}
 s_{23}= Y_{a } Z_{a }~,\quad
s_{223}=d_{abc } Y_{a } Y_{b } Z_{c }~, \quad
s_{233}=d_{abc } Y_{a } Z_{b } Z_{c }~, \quad
 b_{11} [Y,Z]=f_{abc}\, f_{ade}\, Y_b\, Z_c\, Y_d\, Z_e~,
\end{eqnarray}
the y-z-solutions are of the form
\begin{eqnarray}
\!\!\!\!\!\!\!\!\!\!\!\!\!\!\!
\Phi^{YZ}_{n_1,...,n_8}[Y,Z] \!\!\!\!\! & =& \!\!\!\!\!
  {2\over 5} {\omega^4 \over \sqrt{11} \pi^{  3/2}}
 p^{(2)}_{n_1,...,n_8}\left(\omega s_{22},\omega s_{33},\omega s_{23},\omega^{3/2} s_{222},\omega^{3/2} s_{333},
\omega^{3/2} s_{223},\omega^{3/2} s_{233},\omega^{2}b_{11}\right)
\nonumber\\
&&\quad\quad\quad\quad\quad\quad
\times\exp{[-\omega (s_{22}+s_{33})/2 ]}
\end{eqnarray}
and energy
\begin{eqnarray}
\epsilon^{(2)}_{n_1,...,n_8}:=\left[8+2(n_1+n_2)+3(n_3+n_4+n_5+n_6+n_7)+4n_8\right]\omega
\end{eqnarray} 
where $ p^{(2)}_{n_1,...,n_8}$ are the same polynomials as for the x-z case.
After orthogonalisation of those degenerate in energy) they form an ONB with respect to the measure
\begin{equation}
\langle \Phi^\prime_{YZ} |O|\Phi_{YZ}\rangle= 
\int d\mu_Y \int d\mu_Z
\  \  \Phi^\prime_{YZ}  \, O[Y,Z]\, \Phi_{YZ} ~.
\end{equation}

\subsubsection{Trigonal form of the two-direction Hamiltonian}
The polynomials 
\begin{equation}
  p_{n_1,...,n_8}^{(2)}(x_{11},x_{33},x_{13},x_{111},x_{333},x_{113},x_{133},b_{22}) 
\end{equation}
can be easily obtained using the equation
\begin{eqnarray}
\label{D-2}
\Big[D^{(1)}_1+D^{(1)}_3
+ D^{(2)}_{13}\Big] 
 p^{(2)} 
=(\epsilon_{13}/\omega)\  p^{(2)}
\end{eqnarray}
with the one-direction differential operators
\begin{equation}
D^{(1)}_1:= D^{(1)}[x_{11},x_{111}]~, \quad\quad\quad D^{(1)}_3:=  D^{(1)}[x_{33},x_{333}]
\label{D2}
\end{equation} 
and
\begin{equation}
 D^{(2)}_{13}:= D^{(2)}[x_{11},x_{33},x_{13},x_{111},x_{333},x_{113},x_{133},b_{22}]~, 
\end{equation} 
with the two-direction operator $D^{(2)}$ shown explicitly in Appendix D. 
As for the one-direction operators $ D^{(1)}$, the two-direction
operator  $D^{(2)}$ contains an A-power conserving part $D^{(2)}_0$ and a the A-power by two lowering operator $D^{(2)}_{-2}$.
Hence, the eigenvalue problem (\ref{D-2}) is triangular in the space of monomials $M(x_1,...,x_8)$ in $x_1,...,x_8$ and
 can easily be diagonalised with eigenvalues
\begin{eqnarray}
\epsilon=\left(8+2(n_1+n_2)+3(n_3+n_4+n_5+n_6+n_7)+4n_8\right)\omega\equiv \left(8+n\right)\omega
\end{eqnarray}
The eigenfunctions are polynomials
\begin{eqnarray}
p^{(2)}_{n_1,...,n_8}={\cal N}\left(
\sum_{m_1,...,m_2}^{m < n}
\!\!\!\!\ \left[ {\cal A}^{(1)}_{n_1,...,n_8 }(m_1,...,m_8)\ x_1^{m_1}\cdots x_8^{m_8}\right]+x_1^{n_1}\cdots x_8^{n_8}\right)
\end{eqnarray}
with the leading "defining" monomial   $\prod_{i=1}^8 x_i^{n_i}$ of maximal order and 
a "tail" of monomials of decreasing, lower powers 
with some definite coefficients ${\cal A}^{(1)}_{n_1,...,n_8 }(m_1,...,m_8)$.

\subsubsection{Some examples of solutions}
Together, we have
 \begin{equation}
 H_{h.o.}\Phi_{X|YZ}= \left( \epsilon_X+ \epsilon_{YZ}\right) \Phi_{X|YZ} =
\epsilon_{h.o.}\, \Phi_{X|YZ}~, \quad {\rm and\ cycl.\ perm.} 
\end{equation}
with the energy eigenvalues
\begin{eqnarray}
\epsilon_{h.o.}=\left(12+n_X+n_{YZ}\right)\omega=\left(12+n\right)~,\omega\quad {\rm and\ cycl.\ perm.} 
\end{eqnarray}
The lowest solutions are
$
\Phi_{[n]\, X|YZ}[A]= P_{[n]\, X|YZ}[A]\exp[-\omega\left(A_{ai}\right)^2/2]
$
and  cycl. perm., with
\begin{eqnarray}
\epsilon_{h.o.} = 14\,\omega~: &&\!\!\!\!\!\!
P_{[2]\, X|YZ}\propto p^{(1)}_{0, 0}[X]\, p^{(2)}_{0, 0,1,0,0,0,0,0}[Y,Z]\propto \omega s_{23}~,
\quad {\rm and\ cycl.\ perm.}  
\nonumber
\\ 
\epsilon_{h.o.} = 15\,\omega~: &&\!\!\!\!\!\!
P_{[3]\, X|YZ}\propto p^{(1)}_{0, 0}[X]\, p^{(2)}_{0, 0,0,0,0,1,0,0}[Y,Z]\propto \omega^{3/2} s_{223}~,
\quad {\rm and\ cycl.\ perm.} 
\nonumber
\\ 
&&\!\!\!\!\!\!
P_{[3]\, X|YZ}\propto p^{(1)}_{0, 0}[X]\, p^{(2)}_{0, 0,0,0,0,0,1,0}[Y,Z]\propto \omega^{3/2} s_{233}~,
\quad {\rm and\ cycl.\ perm.} 
\nonumber
\\ 
\epsilon_{h.o.} = 16\,\omega~:
&&\!\!\!\!\!\! P_{[4]1\, X|YZ}\propto  p^{(1)}_{0, 0}[X]\, p^{(2)}_{0, 0,0,0,0,0,0,1}[Y,Z]
\propto  6-3\omega ( s_{22}+ s_{33})/2 + \omega^2  b_{11} ~,
\ {\rm and\ cycl.\ perm.}
\nonumber
\\ 
&&\!\!\!\!\!\!
 P_{[4]2\, X|YZ} \propto p^{(1)}_{0, 0}[X]\, p^{(2)}_{0, 0,2,0,0,0,0,0}[Y,Z]
\propto  2- \omega ( s_{22}+ s_{33})/2+ \omega^2  s_{23}^2~,
\ {\rm and\ cycl.\ perm.} 
\nonumber
\\ 
&&\!\!\!\!\!\! P_{[4]3\, X|YZ}\propto  p^{(1)}_{1, 0}[X]\, p^{(2)}_{0, 0,1,0,0,0,0,0}[Y,Z]
\propto(-2+ \omega s_{11}/2)\, \omega  s_{23} ~,
\ {\rm and\ cycl.\ perm.}
\nonumber
\\ 
&&\!\!\!\!\!\! P_{[4]4\, X|YZ}\propto  p^{(1)}_{0, 0}[X]\, p^{(2)}_{1,0,1,0,0,0,0,0}[Y,Z]
\propto  - 5\,\omega s_{23}+\omega^2  s_{22}\, s_{23}~,
\ {\rm and\ perm.}
\nonumber
\\ 
\epsilon_{h.o.} = 17\,\omega~:
&&\!\!\!\!\!\! P_{[5]1\, X|YZ}\propto  p^{(1)}_{1, 0}[X]\, p^{(2)}_{0, 0,0,0,0,1,0,0}[Y,Z]
\propto (-2+\omega  s_{11}/2)\, \omega^{3/2}  s_{223} ~,
\ {\rm and\ cycl.\ perm.}
\nonumber
\\ 
&&\!\!\!\!\!\!
 P_{[5]2\, X|YZ} \propto p^{(1)}_{0, 0}[X]\, p^{(2)}_{1,0,0,0,0,1,0,0}[Y,Z]
\propto - 6 \,\omega^{3/2} \, s_{223}+\omega^{5/2} s_{22}\, s_{223}~,
\ {\rm and\ cycl.\ perm.} 
\nonumber
\\ 
&&\!\!\!\!\!\! P_{[5]3\, X|YZ}\propto  p^{(1)}_{0, 0}[X]\, p^{(2)}_{0,1,0,0,0,1,0,0}[Y,Z]
\propto   - 5 \,\omega^{3/2} \, s_{223}+ \omega^{5/2} s_{33}\, s_{223} ~,
\ {\rm and\ cycl.\ perm.}
\nonumber
\\ ...
\label{X|YZ-sol}
\end{eqnarray}
By superposiition, we obtain from (\ref{X|YZ-sol}) and (\ref{X|Y|Z-sol}),  
all spin-0 solutions up to polynomial order $n=4$, 
namely $\Phi^{(0)++}_{[n]}[A]=P^{(0)++}_{[n]}[A]\exp[-\omega\left(A_{ai}\right)^2/2]$~, with
\begin{eqnarray}
\epsilon^{(0)++}_{h.o.} = 12\,\omega~:
&& 
P_{[0]}^{(0)++}[A]\propto 1~, 
\nonumber\\
\epsilon^{(0)++}_{h.o.} = 14\,\omega~:
&& 
P_{[2]}^{(0)++}[A] \propto  \left[ -12+\omega\, s^{(0)}_{[2]}\right]~,  
\nonumber\\
\epsilon^{(0)++}_{h.o.} = 16\,\omega~:
&&
P_{[4]1}^{(0)++}[A]\propto  \left[ 108 -18\,\omega\,  s^{(0)}_{[2]}
+\,\omega^2  \!  \left(s^{(0)}_{[2]}\right)^2\right] ~,
\nonumber\\
&&
 P_{[4]2}^{(0)++}[A] \propto \left[ 72-12\, \omega  s^{(0)}_{[2]}+\omega^2  s_{[2]ij} s_{[2]ij}\right]~,
\nonumber\\   
&& P_{[4]3}^{(0)++}[A] \propto  \left[  18-3\,\omega\, s^{(0)}_{[2]}+\omega^2 b^{(0)}_{[4]}\right] ~,
\label{all spin-0-sol}
\end{eqnarray}
completing those of (\ref{Spin-0-sol}) up to polynomial order $n=4$ in $A$.

Furthermore, using the notation $s^{(2)}_{[2]\, ij}:= (s_{[2]\, ij}-\delta_{ij} s_{[2]kk}/3)$,
we can build first spin-2 solutions $ \Phi^{(2)++}_{ij}[A]=P^{(2)++}_{ij}[A]\exp[-\omega\left(A_{ak}\right)^2/2]$ with
\begin{eqnarray}
\epsilon_{h.o.} = 14\,\omega~:&&\quad P^{(2)++}_{[2]\, ij}[A]\propto \left[\omega\,  s^{(2)}_{[2]\, ij}\right]~,
\nonumber\\
\epsilon_{h.o.} = 16\,\omega~:&&
\quad P^{(2)++}_{[4]\, ij}[A]\propto \left[-14\,\omega\,  s^{(2)}_{[2]\, ij}+\omega^2  s^{(0)}_{[2]}\,  s^{(2)}_{[2]\, ij}\right]~,
\label{spin-2-sol-a}
\end{eqnarray}
and first vector solutions $ \Phi^{(1)--}_{i}[A]= P^{(1)--}_{i}[A]\exp[-\omega\left(A_{ak}\right)^2/2]$, 
denoting $v_{[3]i}:= s_{[3]ijj}$,
\begin{eqnarray}
\epsilon_{h.o.} = 15\,\omega~:&&\quad P^{(1)--}_{[3]\, i}[A]\propto\left[\omega^{3/2}\, v_{[3]i}\right]~,
\nonumber\\
\epsilon_{h.o.} = 17\,\omega~:&&
\quad P^{(1)--}_{[5]\, i}[A]\propto\left[-15\,\omega^{3/2}\, v_{[3]i}+\,\omega^{5/2}\, s^{(0)}_{[2]}\, v_{[3]i}\right]~.
\label{spin-1-sol-a}
\end{eqnarray}
In order to obtain all solutions, we have to consider the general case of non-separable solutions, discussed in the
next paragraph.

\subsection{Solutions depending on all three directions }

Finally we consider the completely non-saparable case
\begin{equation}
H\Phi[X,Y,Z] = \epsilon\, \Phi[X,Y,Z]~. 
\end{equation}

\subsubsection{Inclusion of the components $s_{123}$ and $(b_{23},b_{13},b_{12})$.}

Including the component $s_{123}$ of the symmetric 3-tensor $s_{[3]\, ijk}[A]$, defined in (\ref{s_ij,s_ijk}),
\begin{eqnarray}
s_{123}&=& d_{abc}X_{a} Y_{b} Z_{c}~,
\nonumber
\end{eqnarray}
depending on all three space directions,
we obtain the further $\epsilon_{h.o.}=15\,\omega$ solution
\begin{eqnarray}
\Phi_{[3]}[X,Y,Z]  & \propto  &
 \Big[ \omega^{3/2}\, s_{123} \Big] \exp{[-\omega (s_{11}+s_{22}+s_{33})/2 ]}~.
\nonumber
\end{eqnarray}
From this, (\ref{X|Y|Z-sol}), (\ref{X|YZ-sol}), we have the spin-3 solution 
$\Phi^{(3)--}_{ ijk}[A]=P^{(3)--}_{ijk}[A]\exp[-\omega\left(A_{ak}\right)^2/2]$ with
\begin{eqnarray}
\epsilon_{h.o.} = 15\,\omega~:&&
\quad P^{(3)--}_{[3]\, ijk}[A]\propto \left[\omega^{3/2}\,  s^{(3)}_{[3]\, ijk}\right]~,
\label{all spin-3-sol}
\end{eqnarray}
with the spin-3 part $s^{(3)}_{[3] ijk}$ of the symmetric 3-tensor $s_{[3] ijk}$, see Appendix E.

Including finally the components $ b_{23}, b_{13}, b_{12}$ of the symmetric 4-tensor $b_{[4]\, ij}[A]$, defined in (\ref{b_ij}),
\begin{eqnarray}
\quad\quad
 b_{23} =f_{abc}\, f_{ade}\, X_b\, Z_c\, X_d\, Y_e~, \quad {\rm and\ cycl.\ perm.}\ \  b_{13},  b_{12}~,
\nonumber
\end{eqnarray}
irreducible in A-space, we obtain the triplet of $\epsilon_{h.o.}=16\,\omega$ solutions
\begin{eqnarray}
\Phi_{[4]}[X,Y,Z]  &  \propto &
 \Big[-3\,\omega s_{23}+
\omega^2 \, b_{23} \Big] \exp{[-\omega (s_{11}+s_{22}+s_{33})/2 ]}~, \quad {\rm and\ cycl.\ perm.}
\nonumber
\end{eqnarray}
Noting also the triplet of $\epsilon_{h.o.}=16\,\omega$ solutions
\begin{eqnarray}
\Phi_{[4]}[X,Y,Z]  & \propto &
 \Big[
\omega s_{23}
+\omega^2 \, s_{12}s_{13} \Big] \exp{[-\omega (s_{11}+s_{22}+s_{33})/2 ]}~, \quad {\rm and\ cycl.\ perm.}
\nonumber
\end{eqnarray}
we obtain all spin-2 solutions $ \Phi^{(2)++}_{[n]ij}[A]=P^{(2)++}_{[n]ij}[A]\exp[-\omega\left(A_{ak}\right)^2/2]$ 
up to $n=4$ with
\begin{eqnarray}
\epsilon_{h.o.} = 14\,\omega~:
&&\quad P^{(2)++}_{[2]\, ij}[A]\propto \left[\omega\,  s^{(2)}_{[2]\, ij}\right]~,
\nonumber\\
\epsilon_{h.o.} = 16\,\omega~:
&&
\quad P^{(2)++}_{[4]\, ij}[A]\propto \left[-14\,\omega\,  s^{(2)}_{[2]\, ij}+\omega^2  s^{(0)}_{[2]}\,  s^{(2)}_{[2]\, ij}\right]~,
\nonumber\\
&&
\quad P^{(2)++}_{[4]\, ij}[A]\propto \left[\omega\,  s^{(2)}_{[2]\, ij}+\omega^2  \left(s_{[2] ik}\,  s_{[2] kj}\right)^{(2)}\right]~,
\nonumber\\
&&
\quad P^{(2)++}_{[4]\, ij}[A]\propto \left[\omega\, s^{(2)}_{[2]\, ij}+ \omega^2\, b^{(2)}_{[4]\, ij}\right]~.
\label{all spin-2-sol}
\end{eqnarray}
completing (\ref{spin-2-sol-a}) up to polynomial order $n=4$ in A.

\subsubsection{Inclusion of the further irrreducible vector $b_{[5]i}^{--}[A]$.}

Similarly, we have the vector solutions $ \Phi^{(1)--}_{i}[A]= P^{(1)--}_{i}[A]\exp[-\omega\left(A_{ak}\right)^2/2]$, with
\begin{eqnarray}
\epsilon_{h.o.} = 15\,\omega~:
&&\quad P^{(1)--}_{[3]\, i}[A]\propto\left[\omega^{3/2}\, v_{[3]i}\right]~,
\nonumber\\
\epsilon_{h.o.} = 17\,\omega~:
&&
\quad P^{(1)--}_{[5]\, i}[A]\propto\left[-15\,\omega^{3/2}\, v_{[3]i}+\,\omega^{5/2}\, s^{(0)}_{[2]}\, v_{[3]i}\right]~,
\nonumber\\
&&\quad P^{(1)--}_{[5]\, i}[A]\propto\left[-8\,\omega^{3/2}\, v_{[3]i}+\omega^{5/2}\, s_{[2] ij}\, v_{[3]j}\right]~,
\nonumber\\
&&\quad P^{(1)--}_{[5]\, i}[A]\propto\left[-9\,\omega^{3/2}\, v_{[3]i}+\omega^{5/2}\, s_{[3]ijk}\, s_{[2] jk}\right]~,
\nonumber\\
&&\quad P^{(1)--}_{[5]\, i}[A]\propto\left[\omega^{5/2}\, b_{[5]i}\right]~,
\label{all spin-1-sol}
\end{eqnarray}
completing (\ref{spin-1-sol-a}) up to polynomial order $n=5$ in $A$. The vector $b_{[5]i}$ in (\ref{all spin-1-sol}) is defined as
\begin{eqnarray}
 b_{[5]i}^{--}[A]\! :=\! d_{abc} B^{\rm hom}_{a i}\! B^{\rm hom}_{b i}\! A_{c i}\!
+\!{1\over 4}\!\left(2s_{jk}s_{123}\! -\! s_{jj}s_{ikk}\! -\! s_{kk}s_{ijj}\right),  \ (i  \neq j \neq k) \!\!~.
\end{eqnarray}
It appears e.g. in the tail of $\epsilon_{h.o.} = 19\,\omega$ solution $ b_{[4] ij}\, v_{[3]j}$    
\begin{eqnarray}
\!\!\!\!\!\!\!\!\!\!\!\!\!\!
 &&\!\!\!\!\!   P^{(1)--}_{[7]\, i}[A]\propto\Big[-{9\over 2}\,\omega^{3/2}\, v_{[3]i}
+\omega^{5/2}\Big(-{5\over 4}\, s_{[2]}^{(0)}\, v_{[3]i}+{3\over 2}\, s_{[2] ij}\, v_{[3]j}-{1\over 4}\, s_{[3]ijk}\, s_{[2] jk}
-5\,  b_{[5]i}\Big)
\nonumber
\\ 
&&\quad\quad\quad\quad \quad\quad \quad\quad  \quad\quad \quad\quad  \quad\quad  \quad\quad \quad\quad
+\omega^{7/2}\, b_{[4] ij}\, v_{[3]j}\Big]~.
\end{eqnarray}
and is irreducible, i.e. not representable as the product of the components of $s_{[2]},s_{[3]}$, and $b_{[4]}$.
Considering the solutions of similar products of components up to $n=10$, we find no further elementary tensors
beyond the four already included, the 6 components of the symmetric 2-tensor $s_{[2]}$, the 10 components of the
symmetric 3-tensor $s_{[3]}$, the 6 components of the further symmetric 2-tensor $b_{[4]}$, and the 3 components of the
vector $b_{[5]}$, not considering the axial sector so far.

\subsubsection{Inclusion of the axial states}

\noindent
Indeed, using the axial scalar 
\begin{eqnarray}
a_{[3]}^{-+}[A] := {1\over 6}\epsilon_{ijk}f_{abc}A_{ai}A_{bj}A_{ck}\equiv f_{abc}X_{a}Y_{b}Z_{c}~,
\end{eqnarray}
we obtain the $\epsilon=15\,\omega$ solution
\begin{eqnarray}
\Phi_{[3]}^{-+}[A] & :=& 
 {\omega^{9/2} \over
 \sqrt{6\pi}}  \ \Big[ \omega^{3/2}\, a_{[3]} \Big] \exp{[-\omega (s_{11}+s_{22}+s_{33})/2 ]}~.
\end{eqnarray}
Now, considering the tail of the $ a_{[3]} v_{[3]i}$  solution with $n=6$,
\begin{eqnarray}
\Phi^{+-}_{[6]i}[A] & \propto & 
 \Big[ -{5\over 2}\omega^{2}\, a_{[4]i}+ \omega^{3}\, a_{[3]} v_{[3]i} \Big] \exp{[-\omega\left(A_{ak}\right)^2/2 ]}~.
\end{eqnarray}
we find that we have to inclued also the axial $n=4$ vector
 the axial vetor
\begin{eqnarray}
 a_{[4]i}^{+-}[A]:=d_{abc}\, B^{\rm hom}_{a i}A_{b i} A_{c i}~,\quad (i=1,2,3)
\end{eqnarray}
which is an irreducible polynomial $\epsilon=16\,\omega$ solution of 4-th order in A.
Furthermore, considering the tails of the $ a_{[3]} b_{[4]}$  solutions with $n=7$,
\begin{eqnarray}
\Phi^{(0)-+}_{[7]}[A] & \propto & 
 \Big[ 30\,\omega^{3/2}\, a_{[3]}+ \omega^{5/2}\left(-4\, a_{[3]} s^{(0)}_{[2]}+3\,a^{(0)}_{[5]}\right)
+ \omega^{7/2}\, a_{[3]}  b^{(0)}_{[4]} \Big] \exp{[-\omega\left(A_{ak}\right)^2/2 ]}~.
\\
\Phi^{(2)-+}_{[7]ij}[A] & \propto & 
 \Big[ \omega^{5/2}\left(2\, a_{[3]} s^{(2)}_{[2]ij}+{3\over 2}\,a^{(2)}_{[5]ij}\right)
+ \omega^{7/2}\, a_{[3] } b^{(2)}_{[4]ij} \Big] \exp{[-\omega\left(A_{ak}\right)^2/2 ]}~.
\end{eqnarray}
we find that we have also to include the symmetric axial 2-tensor
\begin{eqnarray}
a_{[5]ij}^{-+}[A]:=d_{abc}\, B^{\rm hom}_{a k} A_{b k} (d_{cde}\, A_{d i}A_{e j})~,
\quad (i \le j \wedge k\neq i,j)
\end{eqnarray}
into the list of irreducible polynomials.
Finally,  considering the tails of the $ a_{[3]} b_{[5]}$  and $ a_{[4]} b_{[4]}$ solutions with $n=8$,
\begin{eqnarray}
&&\!\!\!\!\!\!\!\!\!\!\!\!\!\!\!\!\!\!
\Phi^{+-}_{[8]i}[A]  \propto  
 \Big[ \omega^{2}\, a_{[4]i}+
\omega^{3}\left(-{1\over 4}\,s^{(0)}_{[2]}\, a_{[4]i}+{1\over 4}\, s_{[2]ij}\, a_{[4]j}-{1\over 2}\, a^{(1)}_{[6]i}\right)
 +\omega^{4}\, a_{[3]} b_{[5]i} \Big] \exp{[-\omega\left(A_{ak}\right)^2\!\! /2 ]}~.
\\
&&\!\!\!\!\!\!\!\!\!\!\!\!\!\!\!\!\!\!
\Phi^{(3)+-}_{[8]ijk}[A]  \propto  
 \Big[ 
\omega^{3}\left({1\over 2}\, a_{[3]}\,s^{(3)}_{[3]ijk}+{7\over 4} \left(a_{[4]i} s_{[2]ij}\right)_{S} ^{(3)} - a^{(3)}_{[6]ijk}\right)
 +\omega^{4} \left(a_{[4]i} b_{[4]ij}\right)_{S} ^{(3)}\Big] \exp{[-\omega\left(A_{ak}\right)^2\!\! /2 ]}~.
\end{eqnarray}
 that we have,
last but not least, to include also the symmetric axial 3-tensor
\begin{eqnarray}
a_{[6]ijk}^{+-}[A]:=
d_{abc}B^{\rm hom}_{a i}B^{\rm hom}_{b j}B^{\rm hom}_{c k}~, \quad (i \le j \le k)~,
\end{eqnarray}
containing spin-1 and spin-3 parts, into the list of irreducible polynomials.
In addition to the four irreducible tensors  $s_{[2]},s_{[3]},b_{[4]}$, and $b_{[5]}$, the axial scalar $ a_{[3]}$, 
the 3 componenst of the
axial vector $a_{[4]}$, the 6 components of the symmetric axial 2-tensor $a_{[5]}$, containing another axial scalar 
and an axial spin-2 part,
and finally the symmetrix axial 3-tensor $a_{[6]}$ have to be included into the list of all irreducible polynomials, in terms of which
all polynomial solutions can be represented. The complete list is shown in Table 1. Their transformation properties 
under spatial rotations are
summarised in Appendix E.
Considering the tails of all solutions from products of components of these eight irreducible symmetric tensors, 
up to maximal order $n=12$,
no further irreducible polynomials have been found.

\begin{table}
$\begin{array}{|c|c|c|}  
\hline
\quad\  {\rm  sym.\ 2\,tensor}\quad\! [2]\!\! & s_{[2]ij}^{++}[A]:=A_{a i}A_{a j}~,\quad (i \le j) &\! 0^{++}\! ,2^{++}\!         
 \\  \hline
\quad\ {\rm  sym.\ 3\,tensor}\quad\! [3]\!\! &s_{[3]ijk}^{--}[A]:=d_{abc}\, A_{a i}A_{b j}A_{c k}~, 
 \quad (i \le j \le k) &\! 1^{--}\! ,3^{--}\!\! 
 \\  \hline
\quad\ {\rm  sym.\ 2\,tensor}\quad\! [4]\!\! & b_{[4]ij}^{++}[A]:=B^{\rm hom}_{a i}B^{\rm hom}_{a j}~,\quad (i \le j) 
  ~,\quad\   B^{\rm hom}_{a i}:= (1/2) \epsilon_{ijk}\, f_{abc}\, A_{b j}A_{c k} &\! 0^{++}\! ,2^{++}\!\!  
\\  \hline
\quad\quad\ \ {\rm vector}\quad\quad\,  [5]\!\!\!\!   &\! b_{[5]i}^{--}[A]\! :=
\! d_{abc} B^{\rm hom}_{a i}\! B^{\rm hom}_{b i}\! A_{c i}\!
+\!{1\over 4}\!\left(2s_{jk}s_{123}\! -\! s_{jj}s_{ikk}\! -\! s_{kk}s_{ijj}\right),   (i \! \neq\! j\! \neq\! k) \!\!  &1^{--}    
 \\  \hline
\quad\ \ {\rm axial\ scalar}\quad\,  [3]\!\!\!  &a_{[3]}^{-+}[A]:=f_{abc}\, A_{a 1}A_{b 2}A_{c 3}
=B^{\rm hom}_{a 1}A_{a 1}=B^{\rm hom}_{a 2}A_{a 2}=B^{\rm hom}_{a 3}A_{a 3} &0^{-+}      
 \\  \hline
\quad\ \ {\rm axial\  vector}\quad\,  [4]\! \! & a_{[4]i}^{+-}[A]:=
d_{abc}\, B^{\rm hom}_{a i}A_{b i} A_{c i}~,\quad (i=1,2,3) &1^{+-}        
 \\  \hline \!\!
{\rm  sym.\, axial\  2\,tensor}\, [5]\!\! & a_{[5]ij}^{-+}[A]:=d_{abc}\, B^{\rm hom}_{a k} A_{b k} (d_{cde}\, A_{d i}A_{e j})~,
\quad (i \le j \wedge k\neq i,j)    &\! 0^{-+}\! ,2^{-+}\!\!
\\  \hline \!\! 
 {\rm  sym.\, axial\  3\,  tensor}\, [6]\!\! &a_{[6]ijk}^{+-}[A]:=
d_{abc}B^{\rm hom}_{a i}B^{\rm hom}_{b j}B^{\rm hom}_{c k}~, \quad (i \le j \le k) &\! 1^{+-}\! ,3^{+-}\!\! 
 \\  \hline
\end{array}$
\caption{Definition of the complete set of eight elementary $SU(3)$-invariant spatial tensors on gauge-reduced A-space. 
The color indices $a,b,c$ are summed over, but the spatial indices $i,j,k$ are not, in all lines of the table. 
Note that for the case $i=j$ in the seventh line one can choose any of the two $ k\neq i,j $, 
both give the same $a_{[5]}$. The second column shows the degree $[n]$ of the tensor (as a polynomial in A).
The last column shows the spin components into which the tensor can be decomposed.}
\end{table}

\subsubsection{Triangular from of the complete Hamiltonian}

We would like to point out here, that as for the 1-dimensional and the 2-dimensional cases, the polynomial solutions 
\begin{eqnarray}
\!\!\!\!\!\! P_{[n]}\left[\omega s_{[2]},\omega^{3/2} s_{[3]},\omega^{2} b_{[4]},\omega^{5/2}b_{[5]},
\omega^{3/2} a_{[3]},\omega^{2} a_{[4]},\omega^{5/2} a_{[5]},\omega^3 a_{[6]}\right] 
\end{eqnarray}
of the general 3-dimensional case, seen as polynomial in the 45 components of the 8 irreducible tsymmeric tensors
can be also obtained using a trigonal differential equation
\begin{eqnarray}
&&\Big[D^{(1)}_1+D^{(1)}_2+D^{(1)}_3
+ D^{(2)}_{23}+ D^{(2)}_{13}+ D^{(2)}_{12}+ D^{(3)}_{123}\Big] 
  P_{[n]}=(12+n)\ P_{[n]} 
\end{eqnarray}
with the three-direction-differential operator $ D^{(3)}_{123}$. As the 1-dim. $D^{(1)}=D_0^{(1)}+D_{-2}^{(1)}$ given in
(\ref{D1}) and the 2-dim.  $D^{(2)}=D_0^{(2)}+D_{-2}^{(2)}$ given in (\ref{D2}) and shown explicitely in Appendix D, the 3-dim 
Differential operator  $D^{(3)} = D^{(3)}_{0}+ D^{(3)}_{-2}$ has a diagonal and a by an power-of-2-lowering part
and hence is diagonalisable analytically. The explicit expression of $D^{(3)}_{123} $ takes several pages and is therefore
not show here explicitly.

\subsubsection{Comment on reducibility of polynomials in original constrained and reduced spaces}

As discussed in the work of Dittner \cite{Dittner}, for $SU(3)$ in original  constrained functional space $\{V_i^{a}, i=1,..,D\}$
there are two  irreducible $SU(3)$-invariant tensors in $D=1$ spatial dimension, 
\begin{eqnarray}
V_1^{a}V_1^{a} \quad {\rm and} \quad d_{abc}V_{a1} V_{b1} V_{c1}~,\quad  (D=1)~,
\end{eqnarray}
nine  irreducible $SU(3)$-invariant tensors for $D=2$, 
\begin{eqnarray}
&&V_1^{a}V_1^{a} ~,\ d_{abc}V^{a}_{1} V^{b}_{1}V^{c}_{1}~,\ V_2^{a}V_2^{a} ~,\  d_{abc}V^{a}_{2} V^{b}_{2}V^{c}_{2}~,
\ V_1^{a}V_2^{a} ~,\  d_{abc}V^{a}_{1} V^{b}_{1}V^{c}_{2}~,\  d_{abc}V^{a}_{1} V^{b}_{2}V^{c}_{2}~,
\nonumber\\
&&\quad\  C^{a}[V] C^{a}[V]~,\ d_{abc}\, C^{a}[V]\, C^{b}[V]\, C^{c}[V]~,
\quad\quad   (D=2)~,
\label{VD2}
\end{eqnarray}
where $ C^{a}[V]:= d_{abc}\, V^b_{1}\, V^c_{2}~.$
The nineth is independent of the first 8, in the sense, 
that it cannot be represented as a sum of products of them. 
It is, however, not primitive because it is related to them via outer products.

\noindent
Furthermore, Dittner proved that for $D\ge 3$ there are maximally 35 independent irreducible $SU(3)$-invariant ("primitive") tensors 
of maximally 6th rank.
For tensors of rank higher than 6, the number of constraints due to outer products exceeds the number of irreducible tensors.

In reduced functional space  $\{A_i^{a}, i=1,..,D\}$, considered here,  we have also two  $SU(3)$-invariant irreducible 
tensors in $D=1$ spatial dimension, 
\begin{eqnarray}
s_{11}=A_1^{a}A_1^{a} \quad {\rm and} \quad s_{111}= d_{abc}A^{a}_{1} A^b_{1} A^c_{1}~,\quad  (D=1)~.
\end{eqnarray}
For reduced functional space in $D=2$ we have different to the constrained case, only 8  $SU(3)$-invariant irreducible tensors
\footnote{the eighth is related to the eighth in (\ref{VD2}) via the identity
$f_{abc}f_{ade}=(2/3)(\delta_{bd}\delta_{ce}-\delta_{be}\delta_{cd})+d_{abd}d_{ace}-d_{abe}d_{acd}$},
\begin{eqnarray}
s_{11}~,s_{111}~,s_{22}~,s_{222}~,s_{12}~,s_{112}~,s_{122}~, \ {\rm and}\ B_3^2~,\quad  (D=2)~.
\end{eqnarray}
In contrast to the case of original constrained functional space, the last 9th tensor, which is of rank 6, is reducible in reduced space 
\begin{eqnarray}
d_{abc} C^{a}[A] C^{b}[A] C^{c}[A]= {1\over 18} s_{12}^3 - {1\over 6} s_{12}\,  s_{11}\,  s_{22}  - {1\over 12} s_{111}\, 
 s_{222} + {3\over 4}  s_{112}\,  s_{122} + 
   {1\over 6} s_{12}\,  B^2_3~,\  (D=2)~.
\end{eqnarray}
For the case of 3-dimensional reduced space here we find eigth irreducible tensors of maximally 6th polynomial order in $A$.
Since the $SU(3)$ gauge is reduced completely, outer products in color space are absent in the reduced approch,
 in contrast to the constrained approach.

\subsection{All solutions of the  corresponding harmonic oscillator problem}

As demonstrated in the preceeding paragraphs, the corresponding harmonic oscillator problem
\begin{equation}
H_{h.o.} (A,P)\, |\Phi_{i,M}^{(J)PC}\rangle =
\epsilon^{(J)PC}_{h.o.}[\omega]\, |\Phi_{i,M}^{(J)PC}\rangle~,
\nonumber
\end{equation}
with the same measure as for the case of Yang-Mills Quantum mechanics (\ref{YM-measure}),
\begin{equation}
\langle\Phi_1|O |\Phi_2\rangle =\int d\mu_X\int d\mu_Y\int d\mu_Z
  \ \Phi_1^{\dagger} O\ \Phi_2 ~,
\label{HO-measure}
\end{equation}
only replacing the chromomagnetic potential by the 16-dim. harmonic oscillator potential with parameter $\omega$, see (\ref{B2toA2}),
turns out to be trigonal in the space of the monomial functionals 

\begin{equation}
 M[\omega s_{[2]},\omega^{3/2} s_{[3]},\omega^{2} b_{[4]},\omega^{5/2}b_{[5]},
\omega^{3/2} a_{[3]},\omega^{2} a_{[4]},\omega^{5/2} a_{[5]},\omega^3 a_{[6]}]\, \exp[- {1\over 2}\omega \left(A_{ai}\right)^2]~,
\end{equation}
and hence integrable.
The $M$ are monomials in the  45 components of eight elementary $SU(3)$-invariant spatial tensors in reduced A-space
shown in Table 1. Note $\left(A_{ai}\right)^2\equiv \left(s_{11}+s_{22}+s_{33}\right)$.

Organising the monomial functionals according to  the degree $n$ (as a polynomial in the $A$) and 
the conserved quantum numbers J,M,P,C
and applying a Gram-Schmidt orthogonalisation with respect to the measure, 
we obtain all exact solutions
\begin{eqnarray}
\Phi_{[n]\,i,M}^{(J)PC}[A]= P_{[n]\,i,M}^{(J)PC}[\omega s_{[2]},\omega^{3/2}s_{[3]},
\omega^{2} b_{[4]},
\omega^{5/2} b_{[5]},
\omega^{3/2} a_{[3]},\omega^{2} a_{[4]},\omega^{5/2} a_{[5]},\omega^3 a_{[6]}]\,
\times \exp[-\omega\left(A_{ai}\right)^2/2]~,
\nonumber
\end{eqnarray}
of the corresponding harmonic oscillator problem with energies
$$\epsilon^{(J)PC}_{h.o.} = \left(12+n\right)\omega~,$$
where $n$ is the degree of $P_{[n]}$ as a polynom in the $A$.

\subsubsection{Lowest order monomials for all symmetry sectors $J^{PC}$ }

The lowest order monomials for each symmetry sector $J^{PC}$ are listed in Appendix F, Tables 6a-6d. In the upper part of Table 6a one
finds the monomials for the case $J^{++}$ for even $J=0,2,4,..$, the first $0_{[n]}^{++}$ and $2_{[n]}^{++}$ monomials up to $n=4$
 can be read of from Equ.(\ref{all spin-0-sol}) and (\ref{all spin-2-sol}) in the last paragraphs. Similarly, the upper half of Table 6b
lists the lowest monomials for the case $J^{--}$ for odd $J=1,3,5,..$, the first $1_{[n]}^{--}$  monomials up to $n=5$ 
and $3_{[n]}^{--}$ monomials up to $n=3$  can be read of from Equ.(\ref{all spin-1-sol}) and (\ref{all spin-3-sol}) respectively.

Next we note that for the sector $2_{[6]}^{++}$ in the upper part of table 6a, there appears the symmetric  part $\left( b_{[4] ik }\ s_{[2] kj}\right)^{(2)}_{\rm sym}$. The corresponding antisymmetric part $\left(\epsilon_{i s t}\ b_{[4] s j }\ s_{[2] j t}\right)$ also of order $n=6$
describes the lowest spin-1 monomial $1_{[6]}^{++}$ shown in the lower part of Table 6a listing the case $J^{++}$ for odd $J=1,3,5,..$ .
Similarly, for the sector $3_{[5]}^{--}$ in the upper part of table 6b, there appears the symmetric parts 
$\left( s_{[3] ijp}\ s_{[2] pk}\right)_{\rm sym} ^{(3)}$ and 
$\left( v_{[3] i}\ s_{[2] j k}\right)_{\rm sym} ^{(3)}$, where the symmetrisation is over the open indices $i,j,k$. 
The corresponding antisymmetric parts $\left( \epsilon_{i s t}\ s_{[3] s j k}\ s_{[2] k t}\right)_{\rm sym} ^{(2)}$
 and $\left(\epsilon_{i s t}\ v_{[3] s }\ s_{[2] j t}\right)_{\rm sym} ^{(2)}$, also of order $n=5$,
describes the lowest spin-2 monomials $2_{[5]}^{--}$ appearing in the lower part of Table 6b listing $J^{--}$ for even  $J=0,2,4,..$.

In the Tables 6a and 6b the number of axial components in a monomial  has to be even, e.g. in the case of the $0_{[6]}^{++}$ state 
$(a^{-+}_{[3]}a^{-+}_{[3]})$. The upper and lower parts of Tables 6c and 6d list the monomials for the case of  $J^{-+}$ and  $J^{+-}$
where an odd number of axial components appear in the monomials.

\subsubsection{Energy spectrum of the harmonic oscillator problem $H_{h.o.}$}

\begin{figure}   
\centering
\epsfig{figure=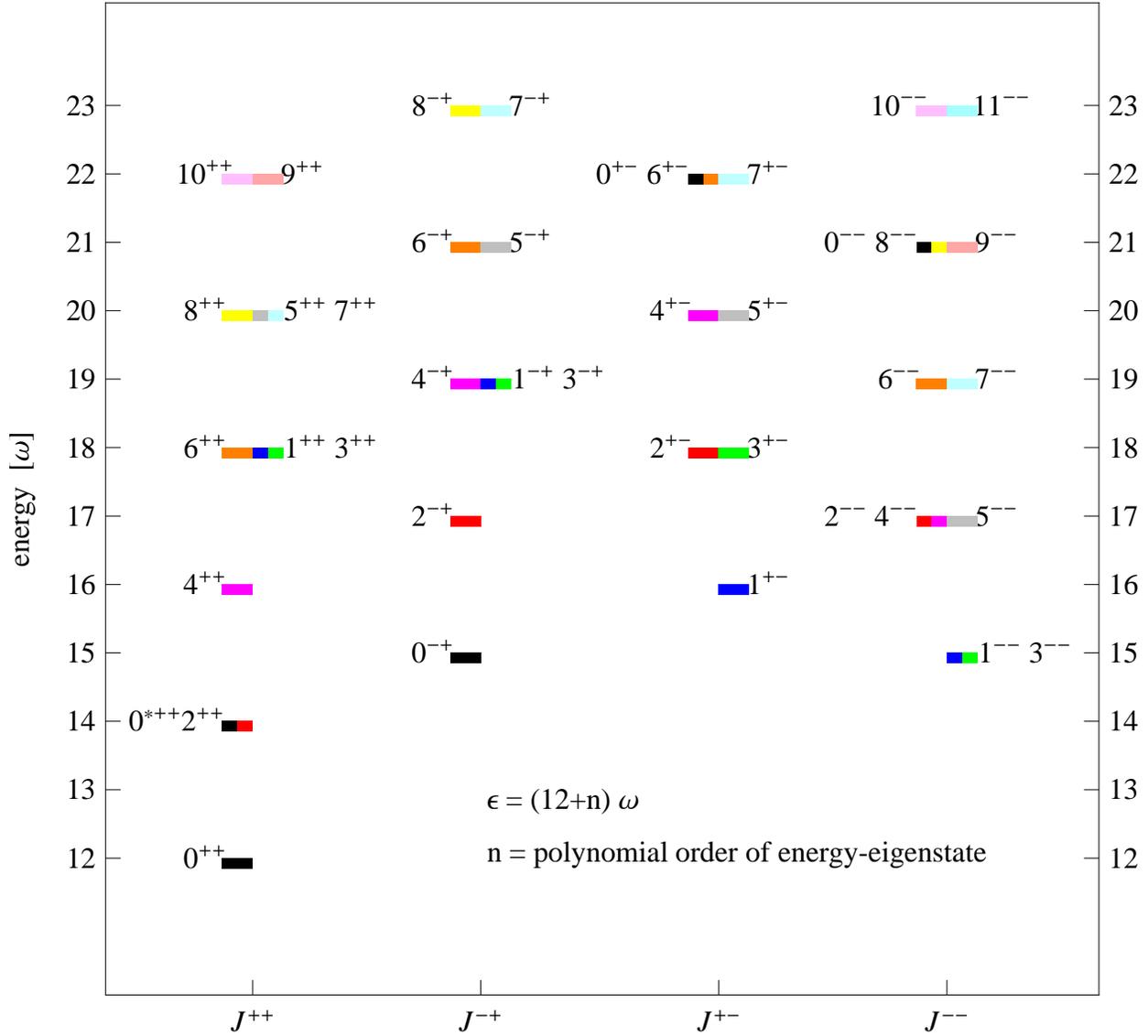,width=17.25cm,angle=0}
\caption
{The lowest energy eigenvalue of the corresponding harmonic oscillator problem $H_{h.o.}$
in each symmetry sector $J^{PC}$ for the (maximal) polynomial order of 10/11 nodes for even/odd parity.} \label{fig1}
\end{figure}

Fig.\ref{fig1} shows the lowest energy eigenvalue of the harmonic oscillator problem $H_{h.o.}$
in each symmetry sector $J^{PC}$ for the (maximal) polynomial order of 10/11 nodes for even/odd parity. 

The spectrum reflects the gauge invariant constructability of the monomials from the components of the irreducible symmetric tensors
for a given symmetry sector. Hence in sectors where the monomials are build from antisymmetric parts of products of tensors,
lie relatively high in energy. For example, the lowest $2^{--}$ state appears only at $n=5$ and therefore energy $\epsilon=19\,\omega$, 
the lowest $1^{++}$ state only at $n=6$ and therefore energy $\epsilon=20\,\omega$. Similarly, their axial colleagues,
 the lowest $2^{+-}$ state appears only at $n=6$ and therefore energy $\epsilon=20\,\omega$, the lowest
$1^{-+}$ states  only at $n=7$ and therefore energy $\epsilon=21\,\omega$.

Very important for the use of a orthogonal basis for the case of SU(3) Yang-Mills Quantum Mechanics is the analytical
construction of the eigensystem of the corresponding harmonic oscillator Hamiltonian $H_{\rm h.o.}$ . This is most effectively done
using Gram-Schmidt orthogonalisation described in the next paragraph.

\subsubsection{Eigenstates of $H_{\rm h.o.}$ from Gram-Schmidt orthogonalisation }

Enumerate all possible monomials for given $J^{PC}$ in increasing order $n$ and multiplicity $m$
\begin{equation}
 M^{(J)PC}_r[A]:=   \omega^{n/2} M_{[n]m}^{(J)PC}\left( s_{[2]}[A], s_{[3]}[A], b_{[4]}[A],b_{[5]}[A],
 a_{[3]}[A],a_{[4]}[A],  a_{[5]}[A],  a_{[6]}[A]\right)~,
\nonumber
\end{equation}
e.g. for the $0^{++}$ sector
\begin{eqnarray}
 M^{(0)++}_1  &=&  M_{[0]}^{(0)++} \equiv 1 ~,
\nonumber\\
 M^{(0)++}_2   &=& \omega\,  M_{[2]}^{(0)++}  \equiv \omega\, s_{[2] i i}~,
\nonumber\\
 M^{(0)++}_{3} &=& \omega^2  \, M_{[4]1}^{(0)++}   \equiv   \omega^2 \,   s_{[2] i i} s_{[2] j j} ~,
\nonumber\\ 
 M^{(0)++}_{4}&=& \omega^2 \,  M_{[4]2}^{(0)++}   \equiv   \omega^2   \,   s_{[2] i j} s_{[2] i j} ~,
\nonumber\\ 
 M^{(0)++}_{5}&=& \omega^2 \,  M_{[4]3}^{(0)++}   \equiv   \omega^2 \,  b_{[4] i i}~.
\nonumber\\
...
\nonumber
\end{eqnarray}
Consider now the Gram matrix    
\begin{eqnarray}
G_{rs}^{(J)PC}:=\langle\langle M_r^{(J)PC}  M_s^{(J)PC} \rangle\rangle_A~,
\label{Gram matrix}
\end{eqnarray}
with the measure 
\begin{eqnarray}
\langle\langle\ \rangle\rangle_A &:=& \int_0^{2\pi} d\psi\, \cos^2[3\,\psi] \int_0^{\infty} d r\, r^{\, 7} \exp[- \omega r^2]
\left[\prod_{a=1}^8\int_{-\infty}^{\infty} d Z_a
\exp[- \omega Z_a^2]\right]
\nonumber\\
&&\!\!\!\!\!\!\!\!\left[\prod_a^{1,2,3,8}\int_{-\infty}^{\infty} d Y_a~.
\exp[- \omega Y_a^2]\right]
\int_{0}^{\infty} d Y_4 Y_4 \exp[- \omega Y_4^2] \int_{0}^{\infty} d Y_6 Y_6  \exp[- \omega Y_6^2]~.
\nonumber
\end{eqnarray}
Here it is very usful that the integration completely factorises into simple 1-dimensional integrations,
which is due to the choice of the flux-tube gauge. Since the functionals to be integrated, are polynomials,
the integrations can be carried out as replacements.

Gram-Schmidt orthogonalisation corresponds to finding a lower-triangular matrix  $T$  such that
$$
 T\, G^{(J)PC} T^T=1~,
$$
obtaining the orthogonal polynomials 
$$  P^{(J)PC}_n[A]:=\sum_{k=1}^{n} T_{nk}M^{(J)PC}_k[A]\quad\quad\quad ~.$$ 
Then the functionals
 $$\Phi^{(J)PC}_n[A,\omega]\equiv P^{(J)PC}_n[A]\ \exp[- {1\over 2}\omega \left(A_{ai}\right)^2]$$
form an ONB of solutions of the corresponding harmonic oscillator problem.
Note that during the orthogonalisation procedure, the linear dependent states appear as zero-eigenvalues of the
Gram matrix $G^{(J)PC}$ Equ.(\ref {Gram matrix}) and can therefore systematically be removed.

The  matrix elements of the harmonic potential can then be easily obtained using
\begin{eqnarray}
 \langle\Phi_m^{(J)PC}[A,\omega]\left( A_{ai}\right)^2\Phi_n^{(J)PC}[A,\omega]\rangle=
 T_{m r}\, T_{n s}\, \langle\langle M_r^{(J)PC}\left( A_{ai}\right)^2\  M_s^{(J)PC}  \rangle\rangle_A
=-{1\over 2}\omega^2{\partial \over \partial \omega}\left(...\right)~.
\label{Vharm}
\end{eqnarray}
Finally, the magnetic matrix elements can be calculated using
\begin{eqnarray}
\langle\Phi_m^{(J)PC}[A,\omega]\left( B_{ai}^2[A]\right)\Phi_n^{(J)PC}[A,\omega]\rangle=
 T_{m r}\, T_{n s}\, \langle\langle M_r^{(J)PC} B_{ai}^2[A]\  M_s^{(J)PC}  \rangle\rangle_A~.
\label{B2harm}
\end{eqnarray}
These will be, as shown in the next section, the main steps to obtain the low energy eigensystem 
of SU(3) YM QM.

\section{ Low-energy spectrum of $SU(3)$ YM QM}

\subsection{The energy spectrum of $SU(3)$ YM QM from the corresponding harmonic oscillator problem }

Consider the  basis of energy eigenstates of the correponding unconstrained harmonic oscillator Schr\"odinger equation 
orthonormal with respect to 
the Yang-Mills measure
\begin{eqnarray}
H_{\rm h.o.}\Phi_n[A,\omega]\equiv \left[T_{\rm kin}+{1\over 2}\omega^2 A_{ai}^2\right]\Phi_n[A,\omega]
=\epsilon^{\rm h.o.}_n \Phi_n[A,\omega]~.\nonumber
\end{eqnarray}
Then the matrix elements of the unconstrained Yang-Mills Hamiltonian are given as
\begin{eqnarray}
{\cal M}_{mn} \!\!\!\!\!\!&:=&\!\!\!\!\!\!
\langle \Phi^\dagger_m[A,\omega]\left(T_{\rm kin}+{1\over 2}  B_{ai}^2[A]\right)\Phi_n[A,\omega]\rangle_A\nonumber\\
\!\!\!\!\!\!&=&\!\!\!\!\!\!\!\!
\left[\delta_{nm}\epsilon^{\rm h.o.}_n-
\langle\Phi^\dagger_m[A,\omega]\!\left({1\over 2} \omega^2 A_{ai}^2\!\right)\!\Phi_n[A,\omega]\rangle_A\right]\!
+{1\over 2}\langle\Phi^\dagger_m[A,\omega]\left( B_{ai}^2[A]\right)\Phi_n[A,\omega]\rangle_A
\nonumber
\end{eqnarray}
since the kinetic terms $T_{\rm kin}$ are the same for the Yang-Mills and the corresponding harmonic oscillator problem.
These are calculated analytically using formulae (\ref{Vharm}) and (\ref{B2harm}).
We treat $\omega$ as a variational parameter, which in each symmetry sector can be choose to minimize the lowest
eigenvalue of the matrix ${\cal M}$, which is easily diagonalised numerically with high accuracy.
The most time consuming part is the calculation of the expectation values of the chromomagnetic potential according
to formula   (\ref{B2harm}), it takes about a week on 36-kernel micro-supercomputers  for truncations 
at polynomial order 10 or 11 for higher spins, and 2-3 months for truncation at polynomial order 12, which is work in
progress.

\subsection{Results}

We have calculated the low energy spectrum of SU(3) Yang-Mills Quantum Mechanics for all symmetry sectors $J^{PC}$
up to spin $J=11$, including polynomials up to 10th order (10-node resolution) for even parity and up to 11th order 
(11-node resolution) for odd parity.

\subsubsection{Lowest level in each symmetry sector $J^{PC}$}

\begin{figure}   
\centering
\epsfig{figure=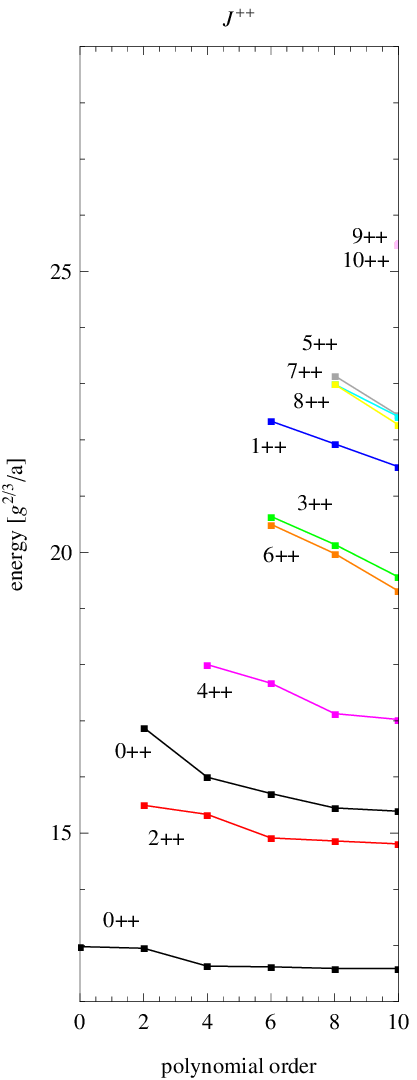,width=4.06cm,angle=0}
\epsfig{figure=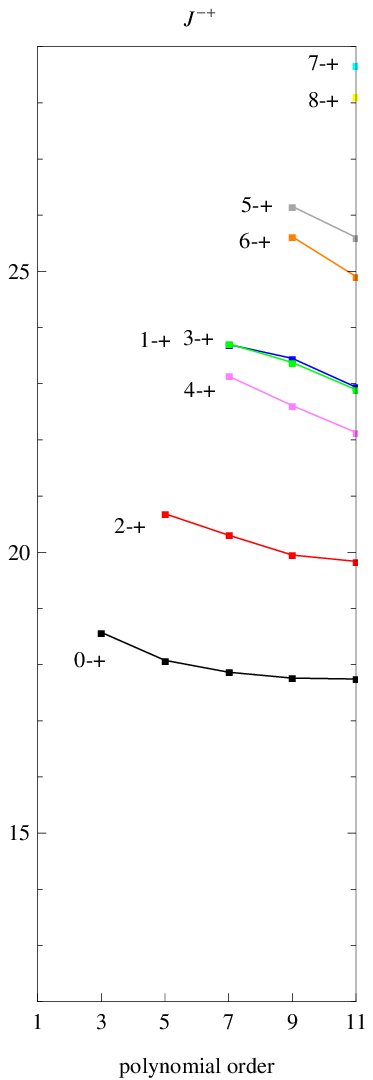,width=4.cm,angle=0}
\epsfig{figure=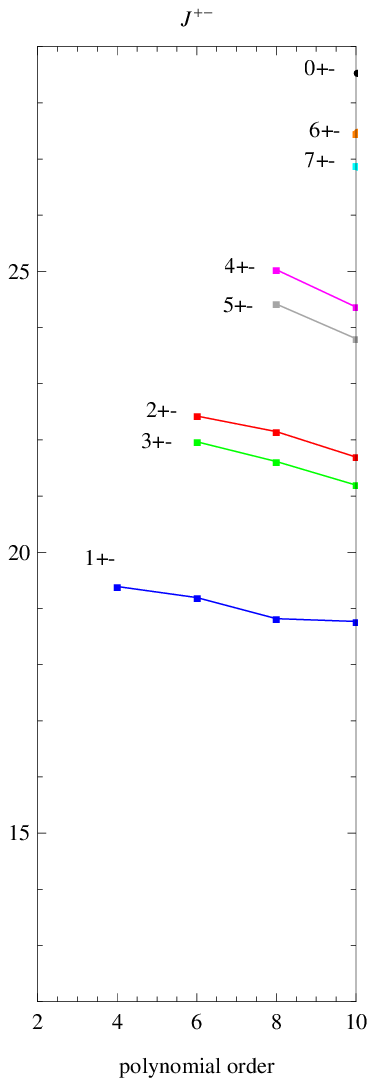,width=4.cm,angle=0}
\epsfig{figure=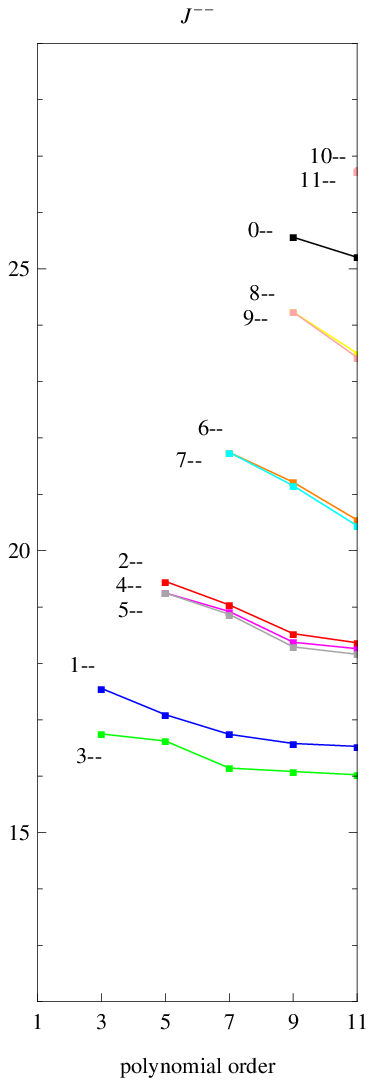,width=4.cm,angle=0}
\caption
{Energy-eigenvalues as a function of polynomial order of truncation. } \label{fig2}
\end{figure}

\begin{figure}   
\centering
\epsfig{figure=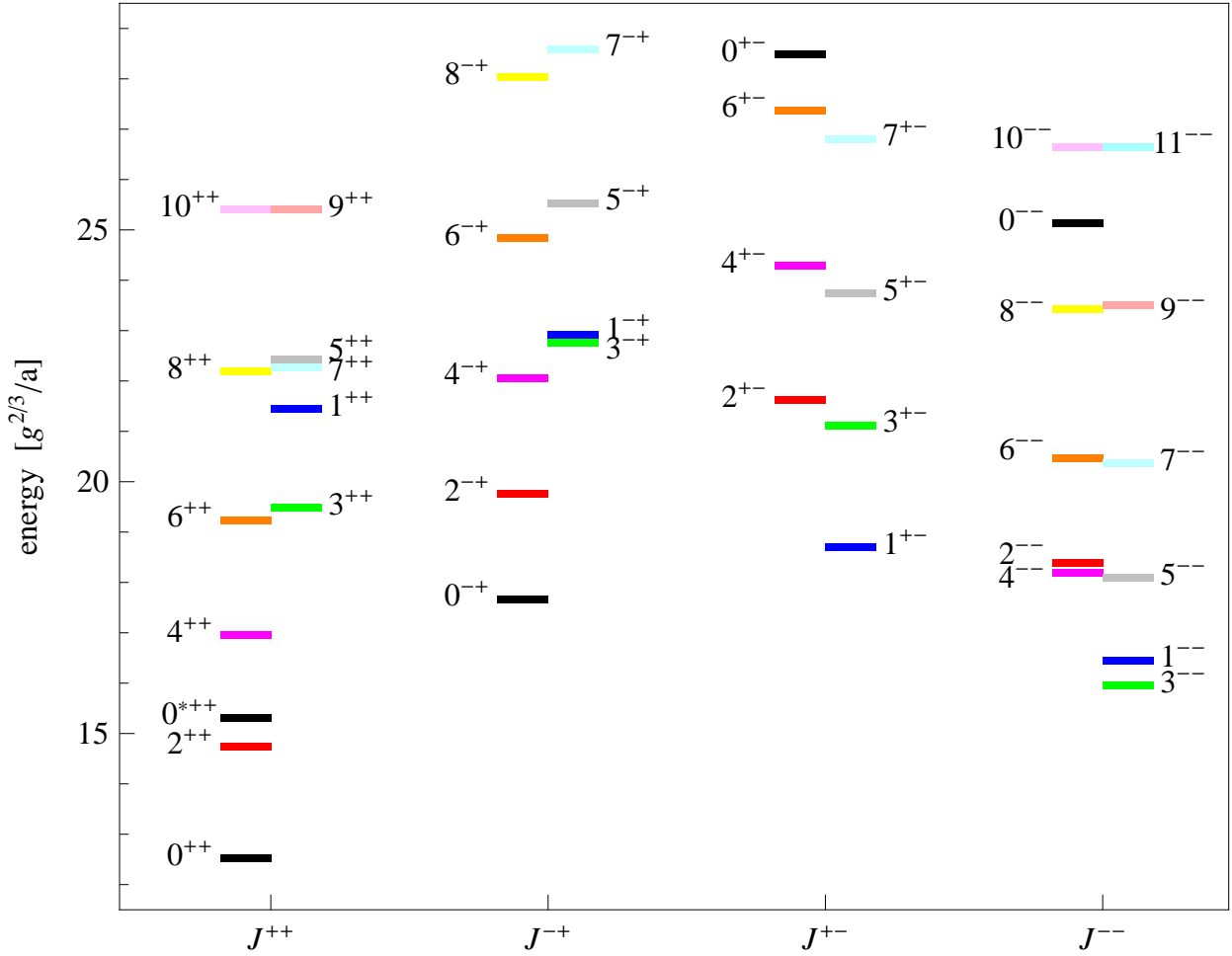,width=17.25cm,angle=0}
\caption
{The lowest energy eigenvalue of SU3 YM-QM in each symmetry sector $J^{PC}$ including all states up to polynomial order
of 10 for even parity and up to polynomial order 11 for odd parity. } \label{fig3}
\end{figure}

Fig.\ref{fig2} shows the lowest energy eigenvalue in each symmetry sector $J^{PC}$ as a function of the polynomial order
of truncation up to 10 for even parity and up to 11 for odd parity. The values are listed in Tables \ref{Tab2a}-\ref{Tab2d}.
Fig.\ref{fig3} shows the lowest energy eigenvalue of SU3 YM-QM in each symmetry sector $J^{PC}$ including all states up to 
polynomial order
of 10 for even parity and up to polynomial order 11 for odd parity. 

The spectrum is purely discrete in accordance with the proof of Simon \cite{Simon} and the groundstate energy is 
obtained to be $\epsilon_{0}^{++}=12.5868~$ (when truncating at 12 nodes).
The lowest states correspond to the $16$ components of $0^{*++}, 2^{++}$ and $1^{--}, 3^{--}$ which show good convergence 
as a function
of increasing polynomial order. These $16$ states correspond to the spins of the
elementary dynamical variables of the "symmetric gauge" \cite{pavel2012}-\cite{pavel2014}.

The higher the polynomial order of truncation, the less the dependence on the variational parameter $\omega$. At 10-th or 11th order of
truncation the results for the spectrum is practically independent of the arbitrarily introduced parameter $\omega$.

Also good convergence as a function of increasing polynomial order show the states $4^{++}$, $0^{-+}$, $5^{--}, 4^{--}, 2^{--}$, 
$1^{+-}$, $2^{-+}$.

For higher states one can use the flow of the energy values with increasing resolution, indicating that at higher energies the
energy levels might be almost equidistant in each symmetry sector as for the corresponding harmonic oscillator spectrum Fig.\ref{fig1}.

Deviations of the spectra of SU(3) Yang-Mills QM Fig.\ref{fig3} with spectra of the corresponding harmonic oscillator Fig.\ref{fig1}
show the effect of the chromomagnetic potential. All aspects and constraints due to gauge invariance are already included in the
corresponding harmonic oscillator spectrum. 

\subsubsection{The lowest few levels in each symmetry-sector $J^{PC}$}

\begin{figure}   
\centering
\epsfig{figure=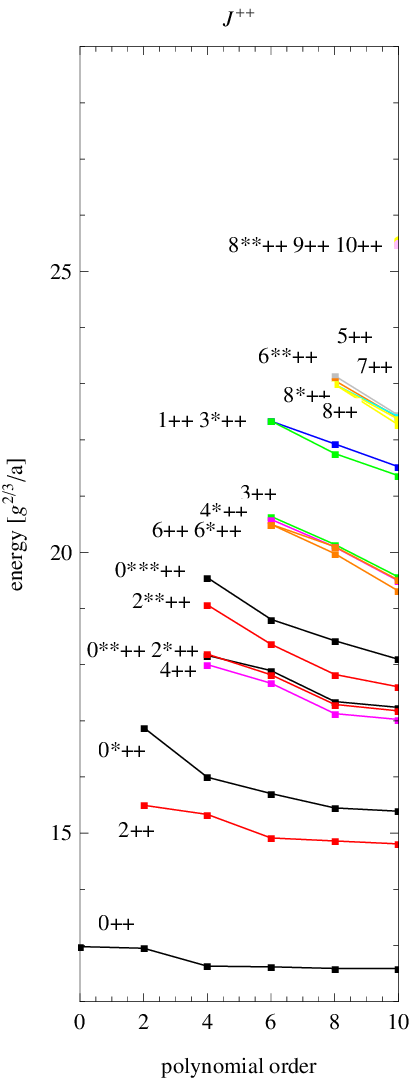,width=4.06cm,angle=0}
\epsfig{figure=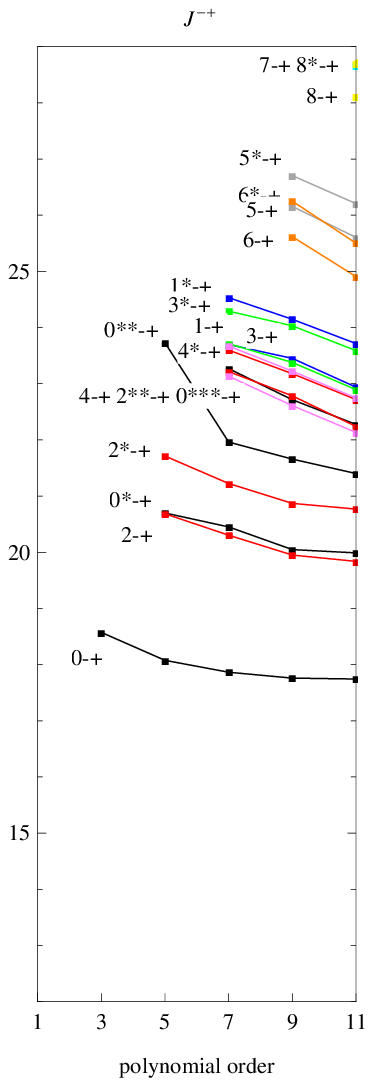,width=4.cm,angle=0}
\epsfig{figure=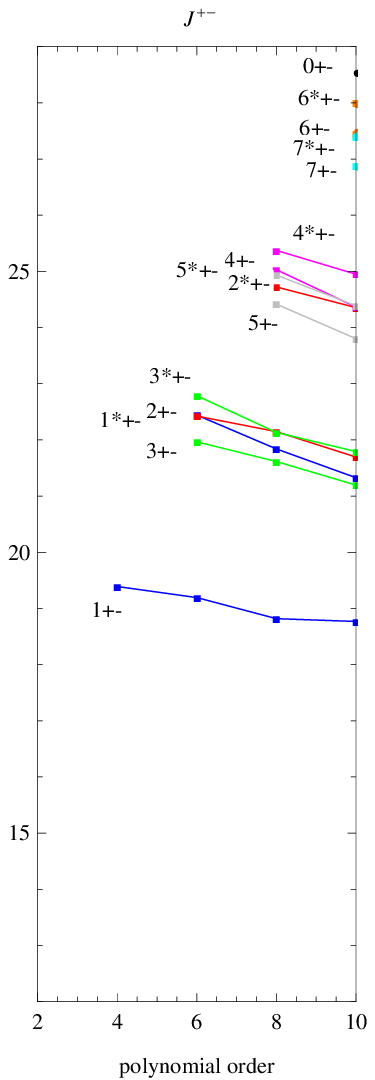,width=4.cm,angle=0}
\epsfig{figure=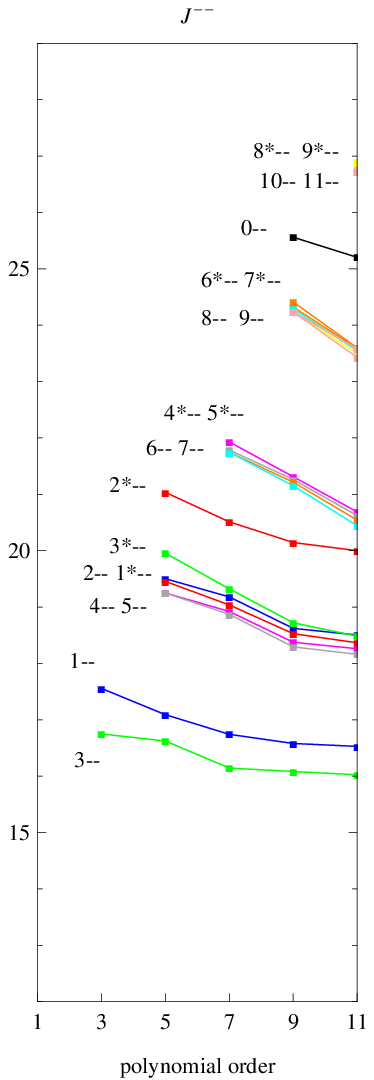,width=4.cm,angle=0}
\caption
{The lowest few energy eigenvalues of SU3 YM-QM in each symmetry sector $J^{PC}$ as a function of the polynomial order
of truncation up to 10/11 nodes for even/odd parity. } \label{fig4}
\end{figure}

Fig.\ref{fig4} shows the lowest few energy eigenvalues of SU3 YM-QM in each symmetry sector $J^{PC}$ as a function of the polynomial order
of truncation up to 10/11 nodes  for even/odd parity. 

Consideration of  the lowest few energy eigenvalues of SU3 YM-QM in each symmetry sector $J^{PC}$ shows, that higher excitations e.g.
the first excited state of spin-2 are close to degenerate to the lowest state of spin-4 similar to the integrable harmonic oscillator problem.

In order to even improve the convergence of the lowest states and to get convergent results also at higher energy, 
polynomials in orders of $14/15$ are necessary. Hence we need very effective
computer algorithms to cope with very large polynomials. This will be subject of future work.

\subsubsection{Comparison with the results of Weisz and Ziemann in the constrained approach}

Theresults found in the present work are in good agreement with the results of Weisz and Zieman \cite{Weisz and Ziemann}, 
shown in the last column denoted by  "WZ" in Tables \ref{Tab2a}-\ref{Tab2d}, using the constrained Hamiltonian approach.
The agreement is excellent in the $0^{++}$ and $2^{++}$, where theire results are already quite accurate.,
although their error estimates turned out to be too optimistic.
Our results are much more accurate values in other sectors considered by them wit only few trial states, 
e.g. in  $1^{--}$ and $3^{--}$ sectors, and we give quite accurate "new results" for the states not considered by them,
as e.g.  $2^{--}, 4^{--}, 5^{--}, 3^{++}.$  

\subsubsection{Comparison with the results of Lattice QCD}

Comparing the results of the low energy spectrum of SU(3) Yang-Mills Quantum Mechanics with those obtained in
Lattice QCD using asymetric lattices, eg. by Morningstar and Peardon \cite{Morningstar} and  Chen et al. \cite{Chen},
using dimensionless results obtained by dividing by the lowest (spin-0) mass,
show reasonable overall agreement for the  $0^{++}, 2^{++}, 3^{++}$,  $0^{-+}, 2^{=+}$, and 
$1^{+-}, 3^{+-}, 2^{+-}, 0^{+-}$ glueball states considered by them. Their $1^{--}, 2^{--}, 3^{--}$ results , however, 
are much higher in energy then those of  Yang-Mills Quantum Mechanics.


\section{Conclusions}

It has been shown in this work, that an unconstrained Hamiltonian formulation of SU(3) Yang-Mills Quantum Mechanics of 
spatially constant fields,
which corresponds to the lowest order in an strong coupling expansion of SU(3) YM theory,  can be carried out in a rather
practical way using the flux-tube gauge. The corresponding Faddeev-Popov operator is simple but non-trivial.
The drawback, that the reduced gauge fields in the fluxtube-gauge themselves are not tensors under spatial rotations,
as was the case for the symmetric gauge on cost of a very complicated FP-operator, 
can be circumvened by forming certain irreducible polynomials of the reduced $A$, eight symmetric tensors, four of which are axial,
which have definite eigenvalues of J,P, and C.

The spectrum of the Hamiltonian of SU(3) Yang-Mills QM of spatially constant fields
can be determined in an effective way using the exact solutions of the corresponding harmonic oscillator problem
only replacing the chromomagnetic potential by a 16-dimensional harmonic oscillator potential parametrised by one 
parameter $\omega$,
but leaving the non-trivial FP-operator unchanged. This model has been demonstrated in this work to be integrable.
The eigensystem turned out to be orthogonal polynomials of the 45 components of the eigth irreducible tensors, four of them axial,
multiplied by a 16-dimensional Gaussian. Its energy spectrum depends only on the polynomial order 
with respect to the $A$ of the eigenstate,
and is highly degenerate. Using the Gram-Schmidt orthogonalisation we could find the eigensystem 
of the corresponding harmonic oscillator
Hamiltonian up to 10th polynomial order for even, and 11th order for odd parity states.
This eigensystem could then be used to find the corresponding eigensystem of SU(3) Yang-Mills Quantum Mechanics
with relatively high accuracy for the low lying states, and the dependence on the variational parameter $\omega$ became very small.
Very helpful for the analytical calculations is here the fact, that uing the flux-tube gauge, the
integrations in functional space for calculating matrix elements completely factorise, and since the eigenstates are polynomials 
multiplied by a 16-dim Gaussian, the integrations can be substituted by replacements.

The results are in good agreement with the results of Weisz and Zieman (1986)
using the constrained Hamiltonian approach in the $0^{++}$ and $2^{++}$ sectors, much more accurate values in other sectors
considered by them, e.g. in  $1^{--}$ and $3^{--}$ sectors , and give quite accurate "new results" for the states not considered by them,
as e.g. $2^{--}$, $3^{++}$ .

By considering the corresponding harmonic oscillator problem, which includes already all effects of gauge invariance, as
an intermediate step, the comparison of its energy spectrum with the final Yang-Mills spectrum shows the effect 
of the chromomagnetic potential.
In order to further investigate the emerging structures in the spectrum, even higher accuracy results and  
polynomials up to order $14/15$ and hence even more effective
programs and algorithms are necessary.
An accurate knowledge of the eigensystem of SU(3) Yang-Mills QM is also necessary for strong coupling perturbation
theory in small $\lambda=g^{-2/3}$ proposed in earlier work \cite{pavel2016}
analogous to the SU(2) approach \cite{pavel2010}.
Analogous to the case of SU(2) Dirac-Yang-Mills QM \cite{pavel2011},
the calculation can also be generalised to the inclusion of quarks to study the masses of mesons.

\section*{Acknowledgements}

I would like to thank A.B. Arbusov, A. Pilloni ,Y. Buistritskij, A. Dorokhov, V. Gerdt, A. Khvedelidze, S. Nedelko, 
F. Niedermayer, Y. Palii, M. Staudacher, O.V. Teryaev, J. Wambach, and P. Weisz
for interesting discussions. This work was partly financed by the SFB 647 "Raum-Zeit-Materie: 
Analytische und Geometrische Structuren."

\begin{appendix}

\section{Explicit form of the inverse of the FP-operator in the flux-tube gauge}

The inverse $\gamma^{-1}$ of the homogeneous part of the Faddeev-Popov operator exists in the regions of non-vanishing
 determinant  (\ref{FP-det}),
and its non-vanishing matrix elements are rather simple,
\begin{eqnarray}
&&\!\!\!\!\!\!\!\!\!\!\!\!
\left(\gamma^{-1}\right)_{12}=-\left(\gamma^{-1}\right)_{21}={1\over r\cos[\psi]}~,  
\nonumber\\
&&\!\!\!\!\!\!\!\!\!\!\!\!
\left(\gamma^{-1}\right)_{32}={1\over 2\, r\cos[\psi]} \left(Y_4/Y_6-Y_6/Y_4\right)~,
\quad 
\left(\gamma^{-1}\right)_{82}=-{1\over 2\sqrt{3}\, r\cos[\psi]} \left(Y_4/Y_6+Y_6/Y_4\right)~,
\nonumber\\
&&\!\!\!\!\!\!\!\!\!\!\!\! ---------------------------------------
\nonumber\\
&&\!\!\!\!\!\!\!\!\!\!\!\!
\left( \gamma^{-1}\right)_{45}=-\left(\gamma^{-1}\right)_{54}=-{1\over r  \cos[\psi+2\pi/ 3]}~,
\quad\quad  
\left(\gamma^{-1}\right)_{34}=-\sqrt{3}\left(\gamma^{-1}\right)_{84}=-{1\over 2\, r  \cos[\psi+2\pi/ 3]} (Y_2/Y_6)~,
\nonumber\\
&&\!\!\!\!\!\!\!\!\!\!\!\!
\left(\gamma^{-1}\right)_{35}={1\over 2\, r  \cos[\psi+2\pi/ 3]} \left(2\,Y_+/Y_4+Y_1/Y_6 \right)~,
\quad 
\left(\gamma^{-1}\right)_{85}={1\over 2\sqrt{3}\, r  \cos[\psi+2\pi/ 3]}\left(2\,Y_+/Y_4-Y_1/Y_6 \right)~,
\nonumber\\
&&\!\!\!\!\!\!\!\!\!\!\!\! ---------------------------------------
\nonumber\\
&&\!\!\!\!\!\!\!\!\!\!\!\!
\left( \gamma^{-1}\right)_{67}=-\left(\gamma^{-1}\right)_{76}={1\over r  \cos[\psi+4\pi/ 3]}~,
\quad\quad  
\left( \gamma^{-1}\right)_{36}=\sqrt{3}\left(\gamma^{-1}\right)_{86}={1\over 2\, r  \cos[\psi+4\pi/ 3]}(Y_2/Y_4)~,
\nonumber\\
&&\!\!\!\!\!\!\!\!\!\!\!\!
 \left( \gamma^{-1}\right)_{37}={1\over 2\, r  \cos[\psi+4\pi/ 3]}\left(Y_1/Y_4-2\,Y_-/Y_6 \right)~,
\quad\quad  
\left(\gamma^{-1}\right)_{87}={1\over 2 \sqrt{3}\, r  \cos[\psi+4\pi/ 3]}\left(Y_1/Y_4+2\,Y_-/Y_6 \right)~,
\nonumber\\
&&\!\!\!\!\!\!\!\!\!\!\!\! ---------------------------------------
\nonumber\\
&&\!\!\!\!\!\!\!\!\!\!\!\!
\left(\gamma^{-1}\right)_{33}=\sqrt{3}\left(\gamma^{-1}\right)_{83}=Y_4^{-1}~,
\quad\quad  
\left(\gamma^{-1}\right)_{38}=-\sqrt{3}\left(\gamma^{-1}\right)_{88}=Y_6^{-1}~,
\nonumber
\end{eqnarray}
grouped into those proportional to $\cos^{-1}[\psi]$, $\cos^{-1}[\psi+2\pi/ 3]$, and $\cos^{-1}[\psi+4\pi/ 3]$, 
and those independent of $r$ and $\psi$.
Such a "Weyl-decomposition" leads to a considerable simplification
of the non-local potential.

\section{Explicit forms of the $I_m^{YZ}\ (m=1,2,4,5,6,7)$ and $I_\pm^{YZ}$}

The explicit expressions of the interations $I_m^{Y\!Z}$ and $I_{\pm}^{Y\!Z}$ in (\ref{H-flux-tube}
) read
\begin{eqnarray}
I_m^{Y\!Z}  &:=&
\left( {1\over Y_4 Y_6}\widetilde{T}_m^{Y\dagger} Y_4\, Y_6+\widetilde{T}_m^{Z}\right)
\left(\widetilde{T}^{Y}_m +\widetilde{T}_m^{Z}\right)
\nonumber\\
I_{\pm}^{Y\!Z} &:=&\left[\left( T^{Z}_3 \pm {1\over \sqrt{3}}T^{Z}_8 \right)+2 T^{Y}_3 
\right]\left( T^{Z}_3 \pm {1\over \sqrt{3}}T^{Z}_8 \right)
\nonumber
\end{eqnarray}
with the shifted non-Hermitean $\widetilde{T}^{Y}_a$  and Hermitean $\widetilde{T}^{Z}_a$, 
defined as
\begin{eqnarray}
\widetilde{T}^{Y}_1 := T^{Y}_1
                               -{1\over 2}\left({Y_6\over  Y_4}
                               -{Y_4\over  Y_6}\right) {T}^{Y}_3~,
\quad &&\quad \widetilde{T}^{Z}_1 := T^{Z}_1
                               -{Y_6\over 2 Y_4}\left( {T}^{Z}_3 +{1\over\sqrt{3} }{T}^{Z}_8\right)
                               +{Y_4\over 2 Y_6}\left( {T}^{Z}_3 -{1\over\sqrt{3} }{T}^{Z}_8\right)~,
\nonumber\\
\widetilde{T}^{Y}_2 :=T^{Y}_2~,\quad\quad\quad\quad\quad\quad\quad\quad\ \  
\quad &&\quad \widetilde{T}^{Z}_2 :=T^{Z}_2~,
\nonumber\\
\widetilde{T}^{Y}_4 := T^{Y}_4
                               -\left({Y_1\over 2 Y_6}
                               +{Y_+\over  Y_4}\right) {T}^{Y}_3 ~,\, 
\quad &&\quad \widetilde{T}^{Z}_4 := T^{Z}_4
                               -{Y_1\over 2 Y_6}\left( {T}^{Z}_3 -{1\over\sqrt{3} }{T}^{Z}_8\right)
                               -{Y_+\over  Y_4}\left( {T}^{Z}_3 +{1\over\sqrt{3} }{T}^{Z}_8\right)~,
\nonumber\\
\widetilde{T}^{Y}_5 := T^{Y}_5
                               -{Y_2\over 2 Y_6} {T}^{Y}_3~,\quad\quad\quad\quad\
\quad &&\quad \widetilde{T}^{Z}_5 := T^{Z}_5
                               -{Y_2\over 2 Y_6}\left( {T}^{Z}_3 -{1\over\sqrt{3} }{T}^{Z}_8\right)~,
\nonumber\\
\widetilde{T}^{Y}_6 := T^{Y}_6
                               +\left({Y_1\over 2 Y_4}
                               -{Y_-\over  Y_6}\right) {T}^{Y}_3~,\, 
\quad &&\quad \widetilde{T}^{Z}_6 := T^{Z}_6
                               +{Y_1\over 2 Y_4}\left( {T}^{Z}_3 +{1\over\sqrt{3} }{T}^{Z}_8\right)
                               -{Y_-\over  Y_6}\left( {T}^{Z}_3 -{1\over\sqrt{3} }{T}^{Z}_8\right)~,
\nonumber\\
\widetilde{T}^{Y}_7 := T^{Y}_7
                               -{Y_2\over 2 Y_4} {T}^{Y}_3 ~,\quad\quad\quad\quad\
\quad &&\quad \widetilde{T}^{Z}_7 := T^{Z}_7
                               -{Y_2\over 2 Y_4}\left( {T}^{Z}_3 +{1\over\sqrt{3} }{T}^{Z}_8\right)~.
\nonumber
\end{eqnarray}

\section{The case of  singular solutions of the X-equation (\ref{X-equation}) }

 The X-equation (\ref{X-equation}) is solved also by the singular functionals
\begin{equation}
\Phi^{\rm sing}_{n_1,n_2}[r, \psi] := 
 { 2  \omega^2 \over
  \sqrt{6} \pi^{3/4} ( \omega\, r^2)^{3/4}} \sqrt{{1 + \sin[3 \psi]\over \cos^2[3 \psi] }}
\ p^{\rm sing}_{n_1,n_2}\left( \omega\, r^2, \sin[3 \psi]\right) e^{- \omega r^2/2 }
\nonumber
\end{equation}
with the energy eigenvalues 
\begin{equation}
\epsilon^{\rm sing}_{\nu, \mu}=\left(5/2+2 n_1+3 n_2\right)\omega=\left(5/2+n\right)\omega
\nonumber
\end{equation}
with the lowest polynomials 
$\quad p^{\rm sing}_{0,0}(x,y)=1~,  \quad  p^{\rm sing}_{1,0}(x,y)= { 1 \over \sqrt{10}}(5 - 2 x)~, \quad ... $.

\noindent
Due to the $ \cos^2[3\,\psi] $ factor in the measure (\ref{FP-op}), absent in the Y- and Z-equations, they are nevertheless finite
and normalisable. 
Hence, when including the regular solutions in other two directions, y and z, we obtain solutions, 
breaking rotational invariance in one direction, the x-direction, leaving only a cylindrical symmetry.


\begin{table}
$$\begin{array}{c|c|c|c|c|c|c|c||c|}
a E  &[0]&\le [2]&\le [4]&\le [6]&\le [8]&\le [10]&\le [12] & WZ\\
\hline
0_1^{++} \!\!  &\!\! 12.980\ (1)\!\! &\!\! 12.951\ (2) \!\! &\!\! 12.632\ (5)\!\!  &\!\!  12.620\ (13)\!\!   &\!\!  12.591\ (30)\!\!  
&\!\! 12.589\ (69)\!\! &\!\! 12.5868 \ (157)\!\! &\!\! 12.5887 \, (106)\!\!\\
\hline
0_2^{++}\!\!   & --  &\!\!   16.879\ (2) \!\!  &\!\! 15.995\ (5)\!\! &\!\! 15.705\ (13) \!\!&\!\!  15.448\ (30) \!\! 
&\!\!  15.390\ (69)\!\! & \!\!  15.351 \ (157)\!\! & 15.38 \ (106)\\
\hline
0_3^{++} \!\!  &  --  & --   &\!\!  18.164 \ (5)\!\!  &\!\! 17.893\ (13)\!\!  &\!\!  17.342 \ (30)\!\!  &\!\! 17.238\ (69)\!\! 
&\!\!   17.137 \ (157)\!\! & 17.23 \ (106)\\
\hline
0_4^{++} \!\!  & --   &  --  &\!\! 19.548 \ (5) \!\!  &\!\! 18.806\ (13) \!\! &\!\!  18.425 \ (30)\!\!  &\!\!  18.095\ (69)\!\! 
&\!\!   17.989 \ (157)\!\!  & --\\
\hline
2_1^{++} \!\!  & -- &\!\!  15.496\ (1) \!\! &\!\!  15.333\ (4) \!\! &\!\!  14.914 \ (13)\!\!   &\!\! 14.861 \ (42)\!\! 
 &\!\!  14.808\ (122)\!\! &  14.800 \ (328)\!\! &\!\! 14.854 \, (69)\!\!\\
\hline
2_2^{++}\!\!   & --  & --   &\!\!  18.187\ (4)\!\! &\!\!17.819 \ (13) \!\!&\!\!  17.291 \ (42) \!\! &\!\!  17.176 \ (122)\!\! 
&  17.091 \ (328)\!\! & 17.26 \ (69)\\
\hline
2_3^{++} \!\!  &  --  & --   &\!\! 19.063  \ (4)\!\!  &\!\!   18.362\ (13)\!\!  &\!\!   17.825 \ (42)\!\!  &\!\!  17.604\ (122)\!\! 
&  17.506 \ (328)\!\! &  -- \\
\hline
4_1^{++}\!\!   & --  & --   &\!\! 18 \ (1)\!\! &\!\!  17.668\ (6) \!\!&\!\!  17.128 \ (25) \!\! &\!\!  17.023 \ (86)\!\!
 &  {\rm in\, cal.} \, (273)\!\! & 18 \ (3)\\
\hline
4_2^{++} \!\!  &  --  & --   & --  &\!\! 20.585 \ (6)\!\!  &\!\!   20.095 \ (25)\!\!  &\!\!  19.491\ (86)\!\! 
&   {\rm in\, cal.} \, (273)\!\! &  -- \\
\hline
6_1^{++} \!\!  &  --  & --   & --  &\!\!  20.497 \ (2)\!\!  &\!\!   19.977  \ (9)\!\!  &\!\!  19.311 \ (38)\!\! & &  -- \\
\hline
6_2^{++} \!\!  &  --  & --   & --  &\!\!  20.497 \ (2)\!\!  &\!\!   20.101  \ (9)\!\!  &\!\!  19.506 \ (38)\!\! &  &  -- \\
\hline
6_3^{++} \!\!  &  --  & --   & --  & --  &\!\!    23.071 \ (9)\!\!  &\!\! 22.395 \ (38)\!\! &  &  -- \\
\hline
8_1^{++} \!\!  &  --  & --   & --  & --  &\!\!   22.989  \ (2)\!\!  &\!\! 22.269 \ (11)\!\! &  &  -- \\
\hline
8_2^{++} \!\!  &  --  & --   & --  & --  &\!\!   22.989  \ (2)\!\!  &\!\!  22.367\ (11)\!\! &  &  -- \\
\hline
8_3^{++} \!\!  &  --  & --   & --  & --  & --  &\!\! 25.508 \ (11)\!\! &  &  -- \\
\hline
10_1^{++} \!\!  &  --  & --   & --  & --  & --  &\!\! 25.478 \ (2)\!\! &   &  -- \\
\hline
10_1^{++} \!\!  &  --  & --   & --  & --  & --  &\!\! 25.478 \ (2)\!\! &  &  -- \\
\hline
\hline
 1_1^{++}  &  --  & --   & --  &\!\! 22.334\ (1)\!\!  &\!\! 21.929\ (8)\!\!  &\!\! 21.523\ (34)\!\! &  & -- \\
\hline
 3_1^{++}  &  --  & --   & --  &\!\!  20.638\ (3)\!\!  &\!\! 20.139\ (15)\!\!  &\!\! 19.560 \ (61)\!\! &  & --  \\
\hline
 3_2^{++}  &  --  & --   & --  &\!\! 22.337 \ (3)\!\!  &\!\!21.758 \ (15)\!\!  &\!\!  21.364 \ (61)\!\! &  & --  \\
\hline
 5_1^{++}   &  --  & --   & --  & --  &\!\!  23.141\ (6)\!\! &\!\! 22.438 \ (34)\!\! &  & -- \\
\hline
 7_1^{++}   &  --  & --   & --  & --  &\!\!  22.989\ (1) \!\! &\!\! 22.414 \ (10)\!\! && -- \\
\hline
 9_1^{++}   &  --  & --   & --  & --  &  --   &\!\! 25.478\ (1)\!\! &  & -- \\
\hline
\end{array}
$$
\caption
{The lowest few energy eigenvalues of SU3 YM-QM in the symmetry sector $J^{++}$ as a function of the polynomial order
of truncation up to 10 nodes. In brackets the number of states. } \label{Tab2a}
\end{table}

\begin{table}
$$\begin{array}{c|c|c|c|c|c||c|}
a E_1  [g^{2/3}] & [3]&\le [5]&\le [7]&\le [9]&\le  [11] & WZ\\
\hline
1^{--}   & 17.555\ (1)  &  17.090\ (5)   &  16.742\ (17)   & 16.580\ (52)  & 16.528\ (143) & 17.05 \ (29)\\
\hline
3^{--}     &  16.749\ (1)  & 16.624\ (4)    &  16.144\ (17)   & 16.083\ (61)  & 16.024\ (193) & 16.5\  (7)\\
\hline
5^{--}   & --   &  19.249\ (1)  & 18.870\ (6)    &  18.293\ (27)   & 18.162\ (104)&  --\\
\hline
7^{--}   &  --  &   --    &  21.744\ (1)  &  21.153\ (8)  &  20.442\ (40)&  -- \\
\hline
9^{--}   &  --  &   --    &   --    &  24.234 \ (2) &  23.429\ (11)& -- \\
\hline
11^{--}   &  --  &   --    &  --    &   --  &  26.722\ (2)& -- \\
\hline
\hline
0^{--} &  --  & --   &  --  & 25.560 \ (3)  &  25.204\ (12) & --  \\
\hline
 2^{--} &  --  & 19.455\ (2)  & 19.037 \ (10)   & 18.528 \ (37) & 18.366  \ (122)  & --  \\
\hline
4^{--}  &  -- &  19.249\ (1)  & 18.921\ (6)    & 18.376 \ (28) &  18.261 \ (107) & --    \\
\hline
6^{--}  &  --  & --   &   21.744\ (1)  &  21.215 \ (10)  &  20.544\ (49)  & --  \\
\hline
8^{--} &  --  & --   &  --  &  24.234\ (1)  &    23.501 \ (13) & -- \\
\hline
10_{1,2}^{--} &  --   & --   &  --  & --  & 26.722 \ (2)  & --  \\
\hline
\end{array}
$$
\caption
{The lowest few energy eigenvalues of SU3 YM-QM in the symmetry sector $J^{--}$ as a function of the polynomial order
of truncation up to 11 nodes. In brackets the number of states.} \label{Tab2b}
\end{table}

\begin{table}
$$\begin{array}{c|c|c|c|c|c||c|}
a E_1  [g^{2/3}] &[3]&\le [5]&\le [7]&\le [9]&\le  [11] & WZ\\
\hline
0^{-+}   &  18.573\ (1)   &  18.076\ (3)   & 17.864\ (9)    & 17.759 \ (26)  & 17.744\ (66) &  17.8 \ (15)\\
\hline
2^{-+}   & --    &  20.685 \ (2)   &  20.304\ (10)   & 19.950\ (37)  & 19.838\ (124) & 21\  (4)\\
\hline
4^{-+}   & --   &    --  & 23.133\ (3)    & 22.610\ (18)  &  22.133\ (81) &  --\\
\hline
6^{-+}   &  --  &   --  &   --  & 25.591\ (5)  &  24.901\ (32)  &  -- \\
\hline
8^{-+}   &  --  &   -- &   --   &   --    & 28.061\ (6) & -- \\
\hline
\hline
1^{-+} & --   &    --  & 23.698\ (2)  & 23.451\ (11)  & 22.939\ (46) & 23\ (4)\\
\hline
3^{-+} & --   &    --   & 23.712\ (2)  & 23.379\ (15)  & 22.895\ (69) & -- \\
\hline
5^{-+} & --   &    --   & --  &  26.098\ (4)  & 25.593 \ (36) & -- \\
\hline
7^{-+} & --   &    --  & --  & -- & 28.589\ (7) & -- \\
\hline
\end{array}
$$
\caption
{The lowest few energy eigenvalues of SU3 YM-QM in the symmetry sector $J^{-+}$ as a function of the polynomial order
of truncation up to 11 nodes. In brackets the number of states.} \label{Tab2c}
\end{table}

\begin{table}
$$\begin{array}{c|c|c|c|c||c|}
a E_1 [g^{2/3}] &[4]&\le [6]&\le [8]&\le [10]&  WZ\\
\hline
 1^{+-}   & 19.390\ (1)   &  19.192\ (5)   & 18.817\ (19)  & 18.768\ (59)& 18 (\rightarrow 19 \, ?) \ (4) \\
\hline
 3^{+-}      & --  &  21.862 \ (3)   & 21.573\ (16)  & 21.144 \ (68)  & -- \\
\hline
 5^{+-}   & --   &  --    & 24.319 \ (4)  &  23.768  \ (26)   &-- \\
\hline
  7^{+-}   & --    & --  &  --  & 26.811  \ (5) &-- \\
\hline
\hline
0^{+-} & -- & --  & --   & 28.552\ (2) & 28\ (4)  \\
\hline
2^{+-} & -- & 22.425\ (1)  & 22.147\ (9) & 21.694\ (43) & 22.1\ (17)   \\
\hline
4^{+-} & --& --    &  24.930\ (4)  & 24.338\ (27) & --    \\
\hline
6^{+-}& --  & --   & -- &  27.428\ (6) & --  \\
\hline
\end{array}
$$
\caption
{The lowest few energy eigenvalues of SU3 YM-QM in the symmetry sector $J^{+-}$ as a function of the polynomial order
of truncation up to 10 nodes. In brackets the number of states. The  $ 1^{+-}$ estimate $'18'$ given in Tab.3 in WZ seems
 to be a misprint, in their Fig.1 it looks more like $'19'$.} \label{Tab2d}
\end{table}

\section{Some explicit expressions for the 2-dim case }

\subsection{Explicit form of the operator $ D^{(2)}=D^{(2)}_{0}+D^{(2)}_{-2}$ }
\begin{eqnarray}
&&
\!\!\!\!\! \!\!\!\!\! 
 D^{(2)}_{0}:=
2\, x_{12}\partial_{ x_{12}}+3\, x_{112}\partial_{ x_{112}}
+3\, x_{122}\partial_{x_{122}}+4\, b_{33}\partial_{ b_{33}}~,
\nonumber\\
 &&
\!\!\!\!\! \!\!\!\!\! 
 D^{(2)}_{-2}:=
-{2\over 3}\left({1\over 4}x_{11}^2+x_{11}x_{22}-b_{33}\right)\partial_{ x_{112}}^2
-{2\over 3}\left({1\over 4}x_{22}^2+x_{11}x_{22}-b_{33}\right) \partial_{ x_{122}}^2
\nonumber\\
&&
-{2\over 3} (x_{11}+x_{22})\, x_{12}  \partial_{ x_{112}} \partial_{ x_{122}}
-2\, x_{12}\left(x_{11} \partial_{ x_{111}} \partial_{ x_{112}}
+x_{22} \partial_{ x_{222}} \partial_{ x_{122}}\right)
\nonumber\\
&&
+2\left({1\over 2}x_{11}x_{22}-b_{33}- x_{12}^2\right)\left( \partial_{ x_{111}} \partial_{ x_{122}}
+ \partial_{ x_{222}} \partial_{ x_{112}}\right)
-{1\over 2}(x_{11}+x_{22})\left(\partial_{ x_{12}}^2+6\, \partial_{ b_{33}}\right)
\nonumber\\
&&
-3\left(x_{112} \partial_{ x_{111}}+x_{122} \partial_{ x_{222}}\right) \partial_{ x_{12}}
-\left((x_{111}+2 x_{122}) \partial_{ x_{112}}+(x_{222}+2 x_{112}) \partial_{ x_{122}}\right)
 \partial_{ x_{12}}
\nonumber\\
&&
-3\left((x_{22} x_{111}+x_{11} x_{122}-2x_{12} x_{112}) \partial_{ x_{111}}
+(x_{11} x_{222}+x_{22} x_{112}-2x_{12} x_{122}) \partial_{ x_{222}}\right)
 \partial_{ b_{33}}
\nonumber\\
&&
-\left((x_{22} x_{112}+x_{11} x_{222}-2x_{12} x_{122}) \partial_{ x_{112}}
+(x_{11} x_{122}+x_{22} x_{111}-2x_{12} x_{112}) \partial_{ x_{122}}\right)
 \partial_{ b_{33}}
\nonumber\\
&&
-{1\over 2}\Big[(x_{11}+x_{22})(-x_{11} x_{22} + x_{12}^2 + 5 b_{33} ) + 
 3 (-(x_{222} x_{112} + x_{111} x_{122}) + (x_{112}^2 + x_{122}^2))
\Big] \partial_{ b_{33}}^2 
\nonumber\\
&&
-\left(2\, x_{12}\partial_{ x_{12}}+4\, b_{33}\partial_{ b_{33}}\right)
\left( \partial_{ x_{11}}+ \partial_{ x_{22}}\right)
-2\, x_{112}\partial_{ x_{112}}\left(2 \partial_{ x_{11}}+ \partial_{ x_{22}}\right)
-2\, x_{122}\partial_{ x_{122}}
\left( \partial_{ x_{11}}+2 \partial_{ x_{22}}\right)~.
\nonumber
\end{eqnarray}

\subsection{Explicit expression for the 2-dim Jacobian in terms of gauge invariant functions}

In the 2-dim case we can perform a coordinate transformation of the eight gauge fields to the eight irreducible gauge invariant polynomials
$$(X_1,X_2,Y_1,Y_2,Y_3,Y_4,Y_6,Y_8)\quad\longrightarrow \quad 
  x:=(x_{11},x_{22},x_{12},x_{111},x_{222},x_{112},x_{122},b_{33})~.$$
Then the matrix elements take the form
\begin{equation}
\langle \Phi^\prime_{XY} |O[X,Y]|\Phi_{XY}\rangle\propto 
\int d^8x
{\exp[- (x_{11}+x_{22})]\over \sqrt{m(x)} }\  p_1^{(2)}(x)  \, O[x]\, p^{(2)}_{2}(x) ~,
\end{equation}
with the polynomial
\begin{eqnarray}
&&  \!\!\!\!\!\!\!\!\!\! \!\! \!\!\!\!\!\!\!\!     
m(x_{11},x_{22},x_{12},x_{111},x_{222},x_{112},x_{122},b_{33})=
\nonumber\\
&&
 (x_{11}^3 - 3 x_{111}^2)(x_{22}^3-3 x_{222}^2)+16\, b_{33}^3 + 36\, b_{33}^2\left(x_{12}^2 - x_{11} x_{22}\right)
+ 12\, b_{33}\Big[2\left(x_{12}^2 - x_{11} x_{22}\right)^2 
\nonumber\\
&&
 +  3\, x_{11}\left( x_{112} x_{222}- x_{122}^2\right)  + 3\,  x_{22}\left( x_{111} x_{122}
 -  x_{112}^2\right)+ 3\, x_{12}\left( x_{112} x_{122}-  x_{111} x_{222}\right) \Big] 
\nonumber\\
&&
+ 4 \left(x_{12}^2 - x_{11} x_{22}\right)^3  -x_{11}^3 x_{22}^3 
 + 9 x_{11}^2 x_{22} \left(x_{122}^2  -2 x_{112} x_{222}\right) + 9 x_{11}  x_{22}^2  \left( x_{112}^2- 2 x_{111} x_{122}\right)
\nonumber\\
&&
+ 12 x_{12}^3 (3 x_{112} x_{122} - x_{111} x_{222})  - 36 x_{12}^2 ( x_{11} x_{122}^2 +  x_{22} x_{112}^2) 
 \nonumber\\
&&
+ 18 x_{12} (x_{11} x_{122} + x_{22} x_{111})  ( x_{11} x_{222}+ x_{22} x_{112}) + 
 \nonumber\\
&&
+ 36 x_{112}^3 x_{222} -54 x_{112} x_{122} x_{111}x_{222}  - 27 x_{112}^2 x_{122}^2 + 36 x_{111} x_{122}^3~.
\end{eqnarray}
The angular momentum operator takes the form
\begin{eqnarray}
J_3&=& -i\Bigg[
2\, x_{12}\left( \partial_{ x_{11}}- \partial_{ x_{22}}\right)
+( x_{22} - x_{11}) \partial_{ x_{12}}
\nonumber\\
&&\quad\quad
+3\left( x_{112}\,\partial_{ x_{111}}- x_{122}\,\partial_{ x_{222}}\right)
+\left(2 x_{122}- x_{111}\right)\partial_{ x_{112}}
+\left(x_{222}-2\, x_{112}\right) \partial_{ x_{122}}\Bigg]~.
\end{eqnarray}


\section{Spin representation of the symmetric tensors}

The simultaneous eigenfunctions of $J_3$ and $J^2\equiv J_1^2+J_2^2+J_3^2$ , given explicitly in (\ref{J-flux-tube}),
for the symmetric 2-tensor $s^{++}_{[2]ij}$  are determined as the spin-0 component
\begin{eqnarray}
s^{(0)++}_{[2]\, 0}= {1\over \sqrt{3}}\left(s_{11}+s_{22}+s_{33}\right)~,
\end{eqnarray}
and the five spin-2  components $s^{(2)++}_{[2]\, ij}:=s^{++}_{[2]\, ij}-\delta_{ ij}s^{++}_{[2]\, kk}/3$~,
\begin{eqnarray}
s^{(2)++}_{[2]\, 0} &=& {1 \over\sqrt{6} }\left(s_{11}+s_{22}-2\,s_{33}\right)~,
\nonumber\\
s^{(2)++}_{[2]\, \pm 1} &=&\pm\, s_{13}+ i\, s_{23}~,
\nonumber\\
s^{(2)++}_{[2]\, \pm 2} &=&- {1\over 2}\left(s_{11}-s_{22}\right)\mp i  s_{12}~.
\end{eqnarray}
For the symmetric 3-tensor $s^{--}_{[3]\, ijk}$ we have
the three spin-1  components $v^{(1)--}_{[3]\, i}:=s^{--}_{[3]\, ijj}$~,
\begin{eqnarray}
v^{(1)--}_{[3]\, 0} &=&\left(s_{113}+s_{223}+s_{333}\right)~,
\nonumber\\
v^{(1)--}_{[3]\, \pm 1} &=&\mp {1\over\sqrt{2}}\left((s_{111}+s_{122}+s_{133})\pm i (s_{112}+s_{222}+s_{233})\right)~,
\end{eqnarray}
as well as the seven spin-3  components  $s^{(3)--}_{[3]\, ijk}:=
s^{--}_{[3]\, ijk}-{1\over 5}\left(\delta_{ jk}v^{--}_{[3]\, i}+\delta_{ ik}v^{--}_{[3]\, j}+\delta_{ ij}v^{--}_{[3]\, k}\right)$~,
\begin{eqnarray}
s^{(3)--}_{[3]\, 0} &=& {1\over\sqrt{30} }\left(3(s_{113}+s_{223})-2\, s_{333}\right)~,
\nonumber\\
s^{(3)--}_{[3]\, \pm 1} &=&\mp {1\over 2\sqrt{10}}\left((s_{111}+s_{122}-4\, s_{133})\pm i (s_{112}+s_{222}-4\, s_{233})\right)~,
\nonumber\\
s^{(3)--}_{[3]\, \pm 2} &=&- {1\over 2}\left(s_{113}-s_{223}\right)\mp  i\, s_{123}~,
\nonumber\\
s^{(3)--}_{[3]\, \pm 3} &=&\pm {1\over 2\sqrt{6}}\left((s_{111}-3\, s_{122})\mp i (s_{222}-3\, s_{112})\right)~.
\end{eqnarray}
Analogous expressions hold for the other irreducible symmteric tensors $b^{++}_{[4]ij}$, $b^{(1)--}_{[5]i}$, as well as the 
irreducible symmetric axial tensors
$a^{(0)-+}_{[3]}$, $a^{(1)+-}_{[4]i}$, $a^{-+}_{[5]ij}$ and $a^{+-}_{[6]ijk}$.

\section{Lowest monomials for all symmetry sectors $J^{PC}$ }

This Appendix presents the Tables 6a-6d showing the monomials of lowest polynomial order in $A$ for each symmetry sector $J^{PC}$.
Here the following conventions are used. Double indices are summed over. The subscript "sym" indicates symmetrization over all open
indices. Furthermore only linearly independent monomials are shown, that is those that are not expressible as sums of
others in the same symmetry sector. For example, the 6th order spin-2 monomial 
\begin{eqnarray}
 \left( s_{[2] i k} s_{[2] k l}  s_{[2] l j}\right)^{(2)}  & \equiv & 
s_{[2] l l} \left( s_{[2] i k}  s_{[2] k j}\right)^{(2)}+(1/2)\left( s_{[2] k l} s_{[2] k l}-s_{[2] k k} s_{[2] l l}\right)  s_{[2] i j}^{(2)}
\end{eqnarray}
is linear dependent and not shown in Table 6a.

$$\begin{array}{|c|l|l}
J^{++}&  M_i^{(J)++} \\
\hline
0^{++}_{[0]}  &  1  \\
\hline
0^{++}_{[2]}   & s^{(0)}_{[2]}\equiv  s_{[2] i i} \\
\hline
0^{++}_{[4]}  &   \left(s^{(0)}_{[2]}\right)^2   ,
\ \ \      s_{[2] i j} s_{[2] i j}  ,
\ \ \    b^{(0)}_{[4]}\equiv b_{[4] i i} \\
\hline
0^{++}_{[6]}  &   \left(s^{(0)}_{[2]}\right)^3    ,
\     s^{(0)}_{[2]}\,  s_{[2] j k} s_{[2] j k}     ,
\       s^{(0)}_{[2]}\, b^{(0)}_{[4]}, 
\           a_{[3]}   a_{[3]}  ,
\        s_{[2] i j} b_{[4] i j}   , 
\     s_{[3] i j k}  s_{[3] i j k},
 \      v_{[3] i}  v_{[3] i}  ,
 \     s_{[2] i j} s_{[2] j k} s_{[2] k i}
        \\
\hline
2^{++}_{[2]}  &  s_{[2] i j}^{(2)}  \\
\hline
2^{++}_{[4]} &      s^{(0)}_{[2]}\,    s_{[2] i j}^{(2)} ,
\       \left( s_{[2] i k} s_{[2] k j}\right) ^{(2)}\!\!\!  ,
\ \       b_{[4] i j}^{(2)}  \\
\hline
2^{++}_{[6]} &      \left(s^{(0)}_{[2]}\right)^2\!  s_{[2] i j}^{(2)} ,\ 
     s^{(0)}_{[2]}  \left( s_{[2] i k} s_{[2] k j}\right) ^{(2)}\!\!\! ,\  \
        s^{(0)}_{[2]}\,  b_{[4] i j}^{(2)} ,\ \
    s_{[2] k l} s_{[2] k l} \,  s_{[2] i j}^{(2)} ,\
    b^{(0)}_{[4]} \, s_{[2] i j}^{(2)} , \\
 &      \left(\! s_{[2] i k}\, b_{[4] k j}\!\right)_{\rm sym} ^{\! (2)} ,\ \
      \left(v_{[3] i}\,  v_{[3] j}\right) ^{\! (2)}\!\!\! ,\ \
    \left(s_{[3] i j k}\,  v_{[3] k}\right) ^{\! (2)}\!\!\! , \ \
   \left(s_{[3] i k l}\,  s_{[3] j k l}\right) ^{\! (2)}
  \\
\hline
4^{++}_{[4]} &   \left(s_{[2] i j}\  s_{[2] k l}\right)_{\rm sym} ^{(4)} \\
\hline
6^{++}_{[6]} &   \left(s_{[2] i j}\  s_{[2] k l}\  s_{[2] m n}\right)_{\rm sym} ^{(6)},
\quad  \left(s_{[3] i j k}\  s_{[3] l m n}\right)_{\rm sym} ^{(6)} \\
\hline
8^{++}_{[8]}  &   \left(s_{[2] i j}\  s_{[2] k l}\  s_{[2] m n}\  s_{[2] s t}\right)_{\rm sym} ^{(8)},
\quad  \left(s_{[3] i j k}\  s_{[3] l m n}\  s_{[2] s t}\right)_{\rm sym} ^{(8)} \\
\hline
10^{++}_{[10]} \! \!    &  
 \left(s_{[2] i j}\  s_{[2] k l}\  s_{[2] m n}\  s_{[2] s t}\  s_{[2] u v}\right)_{\rm sym} ^{(10)},\quad
  \left(s_{[3] i j k}\  s_{[3] l m n}\  s_{[2] s t}\  s_{[2] u v}\right)_{\rm sym} ^{(10)}\\
\hline
\hline
1^{++}_{[6]}  &    \epsilon_{i s t}\ s_{[2]s j }\, b_{[4] j t }   \\
\hline
3^{++}_{[6]} &   \left( \epsilon_{i s t}\ s_{[2] s j}\ b_{[4] t k}\right)_{\rm sym} ^{(3)},
\quad    \left(\epsilon_{i s t}\ s_{[3] s j k}\ v_{[3] t}\right)_{\rm sym} ^{(3)},
\quad   \left(\epsilon_{i s t}\ s_{[2] s j}\ s_{[2] t l}\ s_{[2] l k}\right)_{\rm sym} ^{(3)}  \\
\hline
5^{++}_{[8]}  &   \left(\epsilon_{i s t}\ b_{[4] s j}\  s_{[2] t k}\  s_{[2] l m}\right)_{\rm sym} ^{ \! (5)},
\   \left(\epsilon_{i s t}\ s_{[3] s j k}\  s_{[3] t l p}\  s_{[2] p m}\right)_{\rm sym} ^{ \! (5)}, 
\   \left(\epsilon_{i s t}\ s_{[3] s j k}\  s_{[3] l m p}\  s_{[2] t p}\right)_{\rm sym} ^{ \! (5)}, \\
&   \left( \! \epsilon_{i s t}\ s_{[3] s j p}\  s_{[3] k l p}\  s_{[2] t m} \! \right)_{\rm sym} ^{ \!  (5)} \!  , 
\      \left( \! \epsilon_{i s t}\ s_{[2] s j k}\  v_{[3] t}\  s_{[2] l m} \! \right)_{\rm sym} ^{ \!  (5)} \!  ,
\   \left( \! \epsilon_{i s t}\ s_{[2] s j}\  s_{[2] t p}\  s_{[2] p k}\  s_{[2] l m} \! \right)_{\rm sym} ^{ \! (5)} \\
\hline
7^{++}_{[8]}  &     \left( \epsilon_{i s t}\ s_{[3] s j k}\ s_{[3] l m n}\ s_{[2] r t}\right)_{\rm sym} ^{(7)}   \\
\hline
9^{++}_{[10]}  &     
 \left( \epsilon_{i s t}\ s_{[2] s j}\ s_{[3] t k l}\ s_{[3] m n r}\ s_{[2] u v}\right)_{\rm sym} ^{(9)}~ \\
\hline
\end{array}
$$
Table 6a: Lowest order monomials for the $J^{++}$ sector, for even and odd spin $J$.


$$\begin{array}{|c|l|l}
J^{--} & M^{(J)--} \\
\hline
1^{--}_{[3]}  &   v_{[3] i}\equiv  s_{[3] i j j} \\
\hline
 1^{--}_{[5]}  &   s^{(0)}_{[2]}  v_{[3] i} ,\quad      s_{[2] ij}  v_{[3] j} , 
\quad        s_{[3] i j k} s_{[2] j k},\quad    b_{[5] i}\\
\hline
3^{--}_{[3]}  &   s_{[3] i j k}^{(3)} \\
\hline
3^{--}_{[5]}   &    s^{(0)}_{[2]}  s_{[3] i j k}^{(3)} ,\ \
    \left(s_{[2] i l} s_{[3] l j k}\right)_{\rm sym} ^{(3)} ,\ \
    \left(s_{[2] i j}  v_{[3] k}\right)_{\rm sym} ^{(3)} \\
\hline
5^{--}_{[5]}   &   \left( s_{[3] i j k}\ s_{[2] l m}\right)_{\rm sym} ^{(5)}  \\
\hline
7^{--}_{[7]}  &    \left( s_{[3] i j k}\ s_{[2] l m}\ s_{[2] n s}\right)_{\rm sym} ^{(7)} \\
\hline
9^{--}_{[9]}   &  \left( s_{[3] i j k}\ s_{[2] l m}\ s_{[2] n s}\ s_{[2] t u}\right)_{\rm sym} ^{(9)}, \ \
 \left( s_{[3] i j k}\ s_{[3] l m n}\ s_{[3] s t u}\right)_{\rm sym} ^{(9)},   \\
\hline
11^{--}_{[11]}  &   \left( s_{[3] i j k}\ s_{[2] l m}\ s_{[2] n s}\ s_{[2] t u}\ s_{[2] v w}\right)_{\rm sym} ^{(11)}, \ \
  \left( s_{[3] i j k}\ s_{[3] l m n}\ s_{[3] s t u}\ s_{[2] v w}\right)_{\rm sym} ^{(11)},   \\
\hline
\hline
0^{--}_{[9]}  &  \epsilon_{i j k}\  b_{[4] i l}\  s_{[3] j l m}\ s_{[2] k m},\ \
   \epsilon_{i j k}\  b_{[4] i l}\  v_{[3] j}\ s_{[2] k l}, \ \
  \epsilon_{i j k}\  s_{[3] i l m}\ s_{[2] j m}\ s_{[2] k n}\ s_{[2] l n}, \\
\hline
2^{--}_{[5]}  & \left( \epsilon_{i s t}\ s_{[3] s j k}\ s_{[2] k t}\right)_{\rm sym} ^{(2)},\ \
   \left(\epsilon_{i s t}\ v_{[3] s}\ s_{[2] j t}\right)_{\rm sym} ^{(2)}  \\
\hline
4^{--}_{[5]}   &    \left(\epsilon_{i s t}\ s_{[3] s j k}\ s_{[2] l t}\right)_{\rm sym} ^{(4)}  \\
\hline
6^{--}_{[7]}  &    \left(\epsilon_{i s t}\ s_{[3] s j k}\ s_{[2] l t}\ s_{[2] m n}\right)_{\rm sym} ^{(6)}  \\
\hline
8^{--}_{[9]}  &      
  \left(\epsilon_{i s t}\ s_{[3] s j k}\ s_{[2] l t}\ s_{[2] m n}\ s_{[2] r u}\right)_{\rm sym} ^{(8)}   \\
\hline
10^{--}_{[11]}  &      
 \left(\epsilon_{i s t}\ s_{[3] s j k}\ s_{[2] l t}\ s_{[2] m n}\ s_{[2] r u}\ s_{[2] v w}\right)_{\rm sym} ^{(10)}, \ \
  \left(\epsilon_{i s t}\ s_{[3] s j k}\ s_{[2] l t}\ s_{[3] m n r}\ s_{[3] u v w}\right)_{\rm sym} ^{(10)}\\
\hline
\end{array}
$$
Table 6b: Lowest order monomials for the $J^{--}$ sector, for odd and even spin $J$.


$$\begin{array}{|c|l|l}
J^{-+}& M^{(J)-+} \\
\hline
 0^{-+}_{[3]} &    a_{[3]}  \\
\hline
 0^{-+}_{[5]} &   s^{(0)}_{[2]}   a_{[3]}  ,
\quad    a^{(0)}_{[5]}\equiv  a_{[5] i i}  \\
\hline
2^{-+}_{[5]}  &    a_{[3]}\ s_{[2] i j},\quad
    a_{[5] i j}  \\
\hline
4^{-+}_{[7]} &     a_{[3]} \left(s_{[2] i j}\  s_{[2] k l}\right)_{\rm sym} ^{(4)},
\     \left(a_{[4] i}\  s_{[3] j k l}\right)_{\rm sym} ^{(4)},  \
    \left(a_{[5] i j}\  s_{[2] k l}\right)_{\rm sym} ^{(4)} \\  
 \hline
6^{-+}_{[9]}  &   a_{[3]} \left(s_{[2] i j}\  s_{[2] k l}\  s_{[2] m n}\right)_{\rm sym} ^{(6)},
\   a_{[3]}  \left(s_{[3] i j k}\  s_{[3] l m n}\right)_{\rm sym} ^{(6)},
\   \left(a_{[4] i }\ s_{[3] j k l}\  s_{[2] m n}\right)_{\rm sym} ^{(6)},  \\
&   \left(a_{[5] i j}\  s_{[2] k l}\  s_{[2] m n}\right)_{\rm sym} ^{(6)},
\    \left(a_{[6] i j k}\  s_{[3] l m n}\right)_{\rm sym} ^{(6)}\\  
\hline
8^{-+}_{[11]}  &   a_{[3]} \! \left(\! s_{[2] i j}\  s_{[2] k l}\  s_{[2] m n}\  s_{[2] s t}\!\right )_{\rm sym} ^{(8)}\!\!  ,\ 
 a_{[3]}\!  \left(\! s_{[3] i j k}\  s_{[3] l m n}\  s_{[2] s t}\!\right )_{\rm sym} ^{(8)}\!\! , \ 
  \left(\! a_{[4] i } s_{[2] j k}\  s_{[3] l m n}\  s_{[2] s t}\!\right )_{\rm sym} ^{(8)}\!\! ,  \\
&  \left(a_{[5] i j}\  s_{[2] k l}\  s_{[2] m n}\  s_{[2] s t}\!\right)_{\rm sym} ^{(8)}\! , \ 
  \left( a_{[5] i j}\  s_{[3] k l m}\  s_{[3] n s t}\!\right)_{\rm sym} ^{(8)}\!\!  ,\ 
  \left(\! a_{[6] i j k}\  s_{[3] l m n}\  s_{[2] s t}\!\right )_{\rm sym} ^{(8)} \\ 
\hline
\hline
1^{-+}_{[7]}  &      \epsilon_{i s t}\ a_{[5] s j }\ s_{[2] j t},
\quad      \epsilon_{i s t}\ a_{[4] s }\ v_{[3]  t} \\
\hline
3^{-+}_{[7]}  &   \left( \epsilon_{i s t}\ a_{[5] s j}\ s_{[2] k t}\right)_{\rm sym} ^{(3)},
\  \left(\epsilon_{i s t}\ a_{[4] s}\ s_{[3] t j k}\right)_{\rm sym} ^{(3)}  \\
\hline
5^{-+}_{[9]} &     \left( \epsilon_{i s t}\ a_{[4] s}\ s_{[3] t j k}\ s_{[2] l m}\right)_{\rm sym} ^{(5)},  \ 
   \left( \epsilon_{i s t}\ a_{[4] s}\ s_{[3] j k l}\ s_{[2] t m}\right)_{\rm sym} ^{(5)}, \
 \left( \epsilon_{i s t}\ a_{[5] s j}\ s_{[2] k t}\ s_{[2] l m}\right)_{\rm sym} ^{(5)}, \  \\
&   \left( \epsilon_{i s t}\  a_{[6] s j k}\ s_{[3] t l m}\right)_{\rm sym} ^{(5)} \\
\hline
7^{-+}_{[11]}  &   
 a_{[3]}\left( \epsilon_{i s t}\ s_{[3] s j k}\ s_{[3] l m n}\ s_{[2] r t}\right)_{\rm sym} ^{(7)},    
  \left( \epsilon_{i s t}\  a_{[4] s}\ s_{[3] t j k}\ s_{[2] l m}\ s_{[2] n r}\right)_{\rm sym} ^{(7)},   
\left( \epsilon_{i s t}\  a_{[4] s}\ s_{[3] j k l}\ s_{[2] t m}\ s_{[2] n r}\right)_{\rm sym} ^{(7)},\\
&  
  \left( \epsilon_{i s t}\ a_{[5] s j }\ s_{[2] k l}\ s_{[2] m n}\ s_{[2] r t}\right)_{\rm sym} ^{(7)}, \   
 \left( \epsilon_{i s t}\ a_{[5] s j }\ s_{[3] t k l}\ s_{[3] m n r}\right)_{\rm sym} ^{(7)},\\
&   
   \left( \epsilon_{i s t}\ a_{[6] s j k}\ s_{[3] l m n}\ s_{[2] r t}\right)_{\rm sym} ^{(7)}, 
 \ \left( \epsilon_{i s t}\ a_{[6] j k l}\ s_{[3] s m n}\ s_{[2] r t}\right)_{\rm sym} ^{(7)}\\
\hline
\end{array}
$$
Table 6c: Lowest order monomials for the $J^{-+}$ sector, for even and odd spin $J$.


$$\begin{array}{|c|l|l}
J^{+-}& M^{(J)+-} \\
\hline
 1^{+-}_{[4]} &     a_{[4] i}  \\
\hline
3^{+-}_{[6]}  &  a_{[3]}\  s_{[3] i j k}^{(3)}~,
\quad  \left( a_{[4] i}\  s_{[2] j k}\right)_{\rm sym} ^{(3)},\    a_{[6] i j k}^{(3)}\\
\hline
5^{+-}_{[8]}  &    a_{[3]} \left( s_{[3] i j k}\ s_{[2] l m}\right)_{\rm sym} ^{(5)},   \quad
  \left( a_{[4] i}\ s_{[2] j k}\ s_{[2] l m}\right)_{\rm sym} ^{(5)},  
\quad    \left(a_{[5] i j}\ s_{[3] k l m}\right)_{\rm sym} ^{(5)},\   \left( a_{[6] i j k}\ s_{[2] l m}\right)_{\rm sym} ^{(5)} \\
\hline
7^{+-}_{[10]}  &   a_{[3]}  \left( s_{[3] i j k}\ s_{[2] l m}\ s_{[2] n s}\right)_{\rm sym} ^{(7)},  
\   \left(a_{[4] i}\  s_{[2] j k}\ s_{[2] l m}\ s_{[2] n s}\right)_{\rm sym} ^{(7)},  
\ \left(a_{[4] i}\  s_{[3] j k l}\ s_{[3] m n s}\right)_{\rm sym} ^{(7)} ,  \\
&     \left(\ a_{[5] i j} s_{[3] k l m}\ s_{[2] n s}\right)_{\rm sym} ^{(7)},\
   \left( a_{[6] i j k}\ s_{[2] l m}\ s_{[2] n s}\right)_{\rm sym} ^{(7)} \\
\hline
\hline
0^{+-}_{[10]}  &   \epsilon_{i s t}\  a_{[5] i l}\  s_{[3] s l m}\ s_{[2] t m},
\quad    \epsilon_{i s t}\  a_{[5] i l}\  v_{[3] s}\ s_{[2] t l}, \quad
  \epsilon_{i s t}\  b_{[4] i l}\  a_{[4] s }\ s_{[2] t l} \\
\hline
2^{+-}_{[6]} &   \left(\epsilon_{i s t}\ a_{[4] s}\ s_{[2] j t}\right)_{\rm sym} ^{(2)}  \\
\hline
4^{+-}_{[8]} &    a_{[3]} \left(\epsilon_{i s t}\ s_{[3] s j k}\ s_{[2] l t}\right)_{\rm sym} ^{(4)}~,
   \left(\epsilon_{i s t}\ a_{[4] s }\ s_{[2] l t}\ s_{[2] k l}\right)_{\rm sym} ^{(4)},
  \left(\epsilon_{i s t}\ a_{[2] s j}\ s_{[3] k l m}\right)_{\rm sym} ^{(4)},  
   \left(\epsilon_{i s t}\ a_{[6] s j k}\ s_{[2] l t}\right)_{\rm sym} ^{(4)}\\  
\hline
6^{+-}_{[10]}  &     
a_{[3]}   \left(\epsilon_{i s t}\ s_{[3] s j k}\ s_{[2] l t}\ s_{[2] m n}\right)_{\rm sym} ^{(6)},
\left(\epsilon_{i s t}\ a_{[4] s}\ s_{[3] t j k}\ s_{[3] l m n}\right)_{\rm sym} ^{(6)},
    \left(\epsilon_{i s t}\ a_{[4] s}\ s_{[2]  t j}\ s_{[2] k l}\ s_{[2] m n}\right)_{\rm sym} ^{(6)} , \\ 
&     \left(\epsilon_{i s t}\ s_{[3] s j k}\ a_{[5] l t}\ s_{[2] m n}\right)_{\rm sym} ^{(6)}, 
      \left(\epsilon_{i s t}\ s_{[3] s j}\ a_{[5] k t}\ s_{[3] l m n}\right)_{\rm sym} ^{(6)},
   \left(\epsilon_{i s t}\ a_{[6] s j k}\ s_{[2] l t}\ s_{[2] m n}\right)_{\rm sym} ^{(6)} \\ 
\hline
\end{array}
$$
Table 6d: Lowest order monomials for the $J^{+-}$ sector, for odd and even spin $J$.



\end{appendix}






\begin{thebibliography}{99}
\bibitem{BasMatSav}
G.Z. Basean, S.G. Matinyan and G.K. Savvidi,
JETP Let. 29 (1979) 587.
%
\bibitem{AsaSav}
H.M. Asatryan and G. K. Savvidy, Phys. Lett. A99 (1983) 290.
%
\bibitem{Simon}
B. Simon, Ann. Phys. 146 (1983) 209.
%
\bibitem{Savvidy}
G. K. Savvidy, Phys. Lett. 159B (1985) 325.
%
\bibitem{Medvedev}
B.V. Medvedev, Theor. Math. Phys. 60 (1985) 782.
%
\bibitem{Simonov}
Yu. Simonov, Sov. J. Nucl. Phys. 41 (1985) 835.
%
\bibitem{Martin} C. Martin and D. Vautherin,
Nucl. Phys. B (Proc. Suppl.) 39 B,C (1995) 231.
%
\bibitem{Luescher}
M. L\"uscher, Nucl. Phys. B219 (1983) 233.
%
\bibitem{Luescher and Muenster}
M. L\"uscher and G. M\"unster, Nucl. Phys. B232 (1984) 445.
%
\bibitem{Koller and van Baal}
J. Koller and P. van Baal, Nucl. Phys. B273 (1986) 387;
Nucl. Phys. B302 (1988) 1;
P. van Baal and J. Koller,
Ann. of Phys. (N.Y.) 174 (1987) 299.
%
\bibitem{Weisz and Ziemann}
P. Weisz and V. Ziemann, Nucl. Phys. B284 (1987) 157.
%
\bibitem{KP1}
A.M. Khvedelidze and H.-P. Pavel,
Phys. Rev. D 59 (1999) 105017.
%
\bibitem{KMPR}
A.M. Khvedelidze, D. M. Mladenov, H.-P. Pavel, and G. R\"opke,
Phys. Rev. D 67 (2003) 105013.
%
\bibitem{pavel2007}
H.-P. Pavel,
Phys. Lett. B 648 (2007) 97-106.
%
\bibitem{pavel2010}
H.-P. Pavel, Phys. Lett. B 685 (2010) 353-364.
%
\bibitem{pavel2011}
H.-P. Pavel, 
Phys. Lett. B 700 (2011) 265-276.
%
\bibitem{Calogero}
F. Calogero, J. Math. Phys. 10 (1969) 2191, 2197;  J. Math. Phys.  12 (1971) 419.
%
\bibitem{pavel2012}
H.-P. Pavel,
{\it Unconstrained Hamiltonian formulation of low-energy SU(3) Yang-Mills quantum theory},
arXiv: 1205.2237v1 [hep-th] (2012).
%
\bibitem{pavel2013}
H.-P. Pavel, 
PoS Confinement X (2012) 071, arXiv: 1303.3763 v1 [hep-th] (2013).
%
\bibitem{pavel2014}
H.-P. Pavel, 
 EPJ Web of Conferences {\bf 71} (2014) 00104 , 
arXiv: 1405.1970v1 [hep-th] (2014).
%
\bibitem{pavel2016}
H.-P. Pavel, "SU(3) Yang-Mills Hamiltonian in the flux-tube gauge: Strong coupling expansion and glueball dynamics." arXiv:1611.06542 [hep-th] .
%

%
\bibitem{Morningstar}
C. J. Morningstar and M.J. Peardon, Phys. Rev. D 60 (1999) 034509 [hep-lat/9901004]. 
%
\bibitem{Chen}
Y. Chen {\it et\ al.}, Phys. Rev. D 73 (2006) 014516 [hep-lat/0510074]. 
\bibitem{Christ and Lee}
N.H. Christ and T.D. Lee, Phys. Rev. D 22 (1980) 939.
\bibitem{Dittner}
P. Dittner, Commun. Math. Phys. 22 (1971) 238; ibidem 27 (1972) 44.
%
\end{thebibliography}
\end{document}